\documentclass[fleqn,usenatbib]{mnras}

\usepackage{newtxtext,newtxmath}

\usepackage[T1]{fontenc}

\DeclareRobustCommand{\VAN}[3]{#2}
\let\VANthebibliography\thebibliography
\def\thebibliography{\DeclareRobustCommand{\VAN}[3]{##3}\VANthebibliography}

\usepackage{graphicx}
\usepackage{amsmath}
\usepackage[english]{babel}
\usepackage{ulem}
\usepackage{xcolor}
\usepackage{bm}
\usepackage{subcaption}
\usepackage{array}
\usepackage{diagbox}

\newcommand{\Bd}{\ensuremath{\beta_{\rm d}}}
\newcommand{\Td}{\ensuremath{\rm{T}_{\rm d}}}
\newcommand{\Bs}{\ensuremath{\beta_{\rm s}}}
\newcommand{\fgb}{\texttt{fgbuster}}
\newcommand{\nside}{\texttt{nside}}

\title[Cleaning Galactic foregrounds with fgbuster]{Cleaning Galactic foregrounds with spatially varying spectral dependence from CMB observations with \fgb}

\author[Arianna Rizzieri et al.]{
Arianna Rizzieri,$^{1}$\thanks{E-mail: arianna.rizzieri@physics.ox.ac.uk}
Clément Leloup,$^{2,3}$
Josquin Errard,$^{4}$
Davide Poletti $^{5}$
\\
$^{1}$Department of Physics, University of Oxford, Denys Wilkinson Building, Keble Road, Oxford OX1 3RH, United Kingdom\\
$^{2}$Kavli Institute for the Physics and Mathematics of the Universe (WPI), The University of Tokyo,\\
Institutes for Advanced Study,The University of Tokyo, Kashiwa, Chiba 277-8583, Japan\\
$^{3}$Center for Data-Driven Discovery, Kavli IPMU (WPI), UTIAS, The University of Tokyo, Kashiwa, Chiba 277-8583, Japan\\
$^{4}$Universit\'{e} de Paris Cit\'{e}, CNRS, Astroparticule et Cosmologie, F-75013 Paris, France\\
$^5$SISSA - Scuola Internazionale Superiore di Studi Avanzati,
Via Bonomea 265, 34136, Trieste, Italy.
}

\date{Accepted XXX. Received YYY; in original form ZZZ}

\pubyear{\the\year{}}

\begin{document}
\label{firstpage}
\pagerange{\pageref{firstpage}--\pageref{lastpage}}
\maketitle

\begin{abstract}  
    In the context of maximum-likelihood parametric component separation for next-generation full-sky CMB polarization experiments, we study the impact of fitting different spectral parameters of Galactic foregrounds in distinct subsets of pixels on the sky, with the goal of optimizing the search for primordial $B$ modes.
    Using both simulations and analytical arguments, we highlight how the post-component separation uncertainty and systematic foreground residuals in the cleaned CMB power spectrum depend on spatial variations in the spectral parameters.
    We show that allowing spectral parameters to vary across subsets of the sky pixels is essential to achieve competitive S/N on the reconstructed CMB after component separation while keeping residual foreground bias under control.
    Although several strategies exist to define pixel subsets for the spectral parameters, each with its advantages and limitations, we show using current foreground simulations in the context of next-generation space-borne missions that there are satisfactory configurations in which both statistical and systematic residuals become negligible.
    The exact magnitude of these residuals, however, depends on the mission’s specific characteristics, especially its frequency coverage and sensitivity.
    We also show that the post-component separation statistical uncertainty is only weakly dependent on the properties of the foregrounds and propose a semi-analytical framework to estimate it. On the contrary, the systematic foreground residuals highly depend on both the properties of the foregrounds and the chosen spatial resolution of the spectral parameters.
\end{abstract}



\section{Introduction}
Detecting and characterizing primordial Cosmic Microwave Background (CMB) $B$ modes, in particular through the measurement of the tensor-to-scalar ratio $r$, is one of the goals of upcoming CMB observatories~\citep[see e.g.][]{kamionkowski2016quest}. 
The primordial $B$-mode signal, if present, is expected to have a very low amplitude (a signal RMS $\lesssim \mathcal{O}(1\,{\rm nK}))$, making its detection extremely challenging. 
It requires a very sensitive instrument and an exquisite control of the correlated noise as well as the instrumental and astrophysical systematic effects, in particular at large angular scales where the sensitivity to the cosmological signal is the largest.
In the frequency range where the CMB blackbody peaks ($\sim 100$~GHz), the polarized signal is dominated by the emission of the interstellar medium, itself primarily composed of thermal dust and synchrotron radiation~\citep{krachmalnicoff2016characterization,ade2016planck,akrami2020planck}. 
Lensing $B$ modes and their associated variance are an additional source of uncertainty for the detection of the tensor-to-scalar ratio $r$~\citep[see e.g.][]{errard2016robust}, although significant at smaller scales.

The extraction of cosmological constraints from CMB data sets requires an accurate characterization, in space and frequency, of astrophysical foregrounds.
These emissions are poorly characterized and likely complex (e.g. spatial variations, frequency decorrelation). 
This complexity may limit current data analysis tools and, consequently, reduce the performance of current and next-generation high-sensitivity CMB observatories in probing primordial $B$ modes, such as the Simons Observatory~\citep{ade2019simons} and LiteBIRD~\citep{litebird2023probing}.

Many component separation techniques have been developed, usually characterized as either parametric (assuming an explicit parametrization for the frequency scaling of foreground emissions~\citep{brandt1994separation,eriksen2006cosmic}) or blind (Internal Linear Combination~\citep{2003ApJS..148...97B, tegmark2003high}, Independent Component Analysis~\citep{2002MNRAS.334...53M}, Minimally Informed~\citep{Leloup:2023vkb,Morshed:2024fow}, etc.). 
A limitation to both approaches comes from the spatial variability of the foreground Spectral Energy Densities (SEDs): dealing with this property imposes finding a compromise between the bias and the statistical uncertainty after foreground cleaning. Whereas too few degrees of freedom in the component separation would increase the former, too many would increase the latter. 
In this paper we explore this trade-off in the specific case of a parametric framework.
While the framework is fully general, we focus the discussion on the case of next-generation space-based experiments, for which, as they target large sky fractions, the foreground frequency scaling spatial variability is expected to be a major issue.

To date, two approaches have been proposed to deal with spatial variations of foreground SEDs.
One, based on \cite{chluba2017rethinking} and \cite{2023A&A...669A...5V}, implemented so far in the cross-angular power spectrum method~\citep{wolz2024simons, azzoni2021minimal} but in principle also applicable in other component separation methods. 
It consists in fitting for flexible SEDs, based on a Taylor expansion around the average scaling law.
The other approach consists instead in performing the component separation independently on different sky pixel subsets (e.g.~for parametric methods~\cite{errard2019characterizing, grumitt2020hierarchical, puglisi2022improved}, and for blind methods~\cite{khatri2019data, carones2023multiclustering}), whose definition depends on the features of each component separation method.
In this work we focus on the second approach for parametric component separation, extending on the work of~\cite{errard2019characterizing}.
We focus here on the framework implementation, both in a simulation-based and a semi-analytical approach, and its response to different pixel subset choices.
Instead, we refer the reader to \cite{kabalan2025} for an application of this same parametric method to a LiteBIRD-like scenario showing an optimized agnostic way to define the pixel subsets.

The first incarnations of map-based parametric methods~\citep[e.g.][]{brandt1994separation, eriksen2006cosmic} fitted a single set of spectral parameters per pixel.
This made them robust to the spatial variability of foreground frequency scalings, but required a downgrading of the resolution in order to suppress statistical noise from having too many parameters. \cite{Stompor:2008sf} introduced a more flexible approach that allowed fitting one spectral parameter across a collection of pixels, which went to the opposite extreme of fitting a single spectral parameter set on the full sky in~\cite{Stompor:2016hhw}.
Subsequent works then explored intermediate approaches, breaking the maps into subsets of pixels~\citep{errard2019characterizing, grumitt2020hierarchical, puglisi2022improved}.
A wide variety of approaches can be employed to construct these spatial domains for parametric methods. 
As discussed in \cite{errard2019characterizing}, the two extremes are (a) to fit for a single set of spectral parameters across the entire sky, or (b) to fit for as many sets of spectral parameters as available sky pixels. 
In the ensemble average of CMB and noise realizations, (a) would lead to an erroneous fit of the spectral parameters, and therefore to a large amplitude of the bias on cosmological parameters such as  $r$, depending on the actual variations present in the sky.
In the case of (b), instead, each sky realization would have large foreground residuals in the recovered CMB map due to the low signal-to-noise ratio (S/N) in each pixel of the frequency maps to reconstruct the spectral parameters.
Our generalized parametric component separation implementation allows for any possible scenario in between (a) and (b), moving the choice of the pixel subsets as a preparatory step, independent of the component separation itself but with significant consequences on the foreground cleaning performance and the nature of the foreground residuals in the recovered CMB.
We discuss how in this framework we can aim for the best exploitation of an instrument's sensitivity (in terms of white noise amplitude, but also lever arm for foreground SED characterization) in the presence of spatially varying spectral properties.

The paper is organized as follows.
In Sect.~\ref{sec:compsep_motivation_intuitions} we motivate the need for a component separation method that is robust against the spatial variability of foreground SEDs, and present some intuitions on how a maximum-likelihood component separation works and its performances.
In Sect.~\ref{sec:sp_lik_based_compsep}, we describe the formalism and implementation of the \texttt{fgbuster} simulation-based and semi-analytical tools.
In Sect.~\ref{sec:illustration}, we apply the component separation method to sky simulations based on the proposed next-generation satellite experiments \textit{LiteBIRD}~\citep{litebird2023probing} and \textit{PICO}~\citep{hanany2019pico}, in a variety of scenarios for the independent pixel subsets for the reconstruction of the spectral parameters and we discuss the statistical and systematic residuals trade-off in the different cases.
We also compare the simulation-based analysis and the semi-analytical forecasting.
Finally, we summarize our conclusions in Sect.~\ref{sec:conclusion}.

\section{Component separation motivation and intuitions}
\label{sec:compsep_motivation_intuitions}
We first motivate the need to address the spatial variability complexity in light of the foreground models present in the literature, and we make a general introduction to standard maximum-likelihood parametric component separation.

\subsection{Motivations from observed astrophysical foregrounds}

\begin{figure*}
    \centering
    \includegraphics[width=14cm]{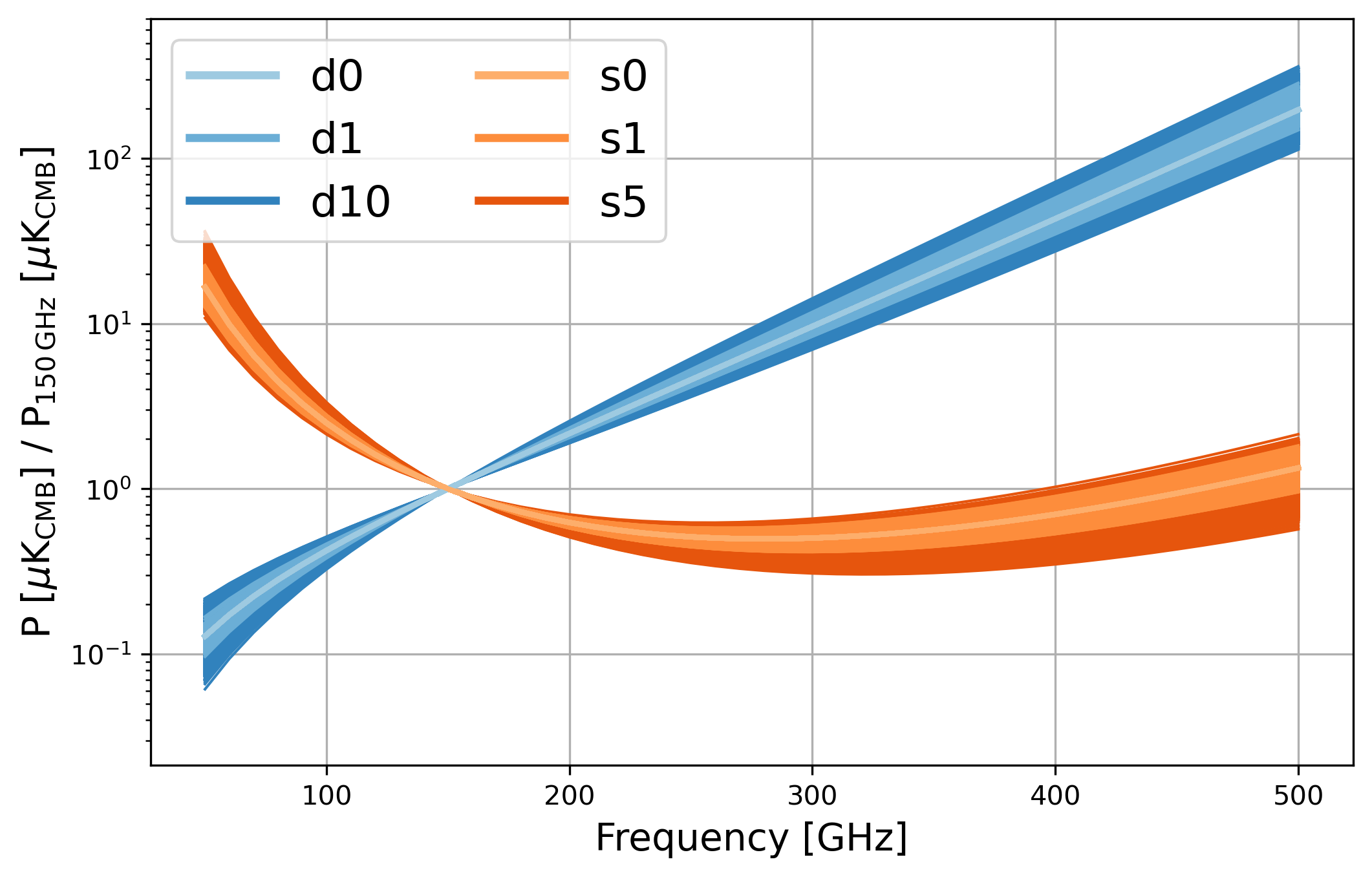}
    \caption{Polarization amplitudes $P\equiv\sqrt{Q^2+U^2}$ between $50$ and $500$~GHz, normalized at $150$~GHz, computed for each pixel (at \texttt{nside}~=~64) for the reference \texttt{PySM3} models \texttt{d0} and \texttt{s0} and the more realistic \texttt{PySM3} models considered in this work \texttt{d1}, \texttt{d10} and \texttt{s1}, \texttt{s5}.
    Each line corresponds to a sky pixel.
    All the lines for the different sky pixels are overlapped for \texttt{d0} (\texttt{s0}), being the same MBB (PL), while for the more realistic models the lines corresponding to the different pixels are spread over a wide range.
    Note that the frequency scalings shown here come from $P$ taken in $\mu \mathrm{K}_{\mathrm{CMB}}$ units, which explains why they do not look like what one could naively expect for a MBB and a PL.
    }
    \label{fig:P_fgs_vs_freq}
\end{figure*}

The latest observations and modeling of the polarized Galactic foregrounds suggest that the dust and synchrotron emission frequency scalings change depending on the position on the sky~\citep{krachmalnicoff2018s, akrami2020planck, delaHoz:2023yvz}.
Fig.~\ref{fig:P_fgs_vs_freq} shows the normalized SED as a function of frequency for three simulated foreground scenarios using \texttt{PySM3}~\citep{Thorne_2017, Zonca_2021, borrill2025full} for thermal dust emission called \texttt{d0}, \texttt{d1} and \texttt{d10}, and for synchrotron emission, called \texttt{s0}, \texttt{s1} and \texttt{s5}. 
In these models, dust emission in each sky pixel $p$ follows a modified blackbody (MBB) emission law,
\begin{equation}
    d^{\text{dust}, X}_{\nu, p} = s_{p}^{\text{d}, X} \left(\frac{\nu}{\nu_0}\right)^{\beta_{\text{d}, p}} \frac{B_\nu\left(\mathrm{T}_{\text{d},p}\right)}{B_{\nu_{0}}\left(\mathrm{T}_{\text{d},p}\right)},
\label{eq:As_dust}
\end{equation}
where $X \equiv \{Q,U\}$, $s^{\text{d}}$ is the dust amplitude at the reference frequency $\nu_{0} = 353$~GHz, $B$ is the black-body spectrum at dust temperature $\Td$ and $\Bd$ the spectral index that parameterizes the frequency dependence of the dust opacity.
The $p$ subscript indicates a dependence on the sky pixels.
These quantities are based on maps derived from Haslam~\citep{haslam1981408, haslam1982408, remazeilles2015improved}, WMAP~\citep{bennett2013nine}, and Planck data~\citep{adam2016planck, akrami2020planck}, smoothed to a resolution $\text{FWHM}=2.6 ^\circ$, and with smaller scales added as Gaussian random fluctuations. 
In the case of \texttt{d0}, the spectral parameters, $\Td$ and $\Bd$, are independent from the position in the sky, and are fixed to $\Bd = 1.54$ and $\Td=20 \, {\rm K}$. 
While in \texttt{d10} additional random small scales are added to the spectral parameter templates.
The synchrotron emission follows a power law (PL) defined as:
\begin{equation}
    d^{\text{sync}, X}_{\nu, p} =  s_{p}^{\text{s}, X} \left(\frac{\nu}{\nu_0}\right)^{\beta_{\text{s},p}}
\label{eq:As_sync}
\end{equation}
where $\Bs$ is the power-law index and $\nu_{0} = 23$~GHz is a reference frequency. 
The $s^\text{s}$ amplitudes and the $\Bs$ templates are derived from Haslam~\citep{haslam1981408, haslam1982408, remazeilles2015improved}, WMAP~\citep{bennett2013nine}, Planck data~\citep{adam2016planck, akrami2020planck} and S-PASS~\citep{carretti2019s} data.
Similarly to the dust case, the amplitude maps are smoothed to a resolution of 5$^\circ$, and smaller scales are added as Gaussian random fluctuations.
In the case of \texttt{s0}, the power-law index is constant across the sky and fixed to $\Bs=-3$.
In the case of \texttt{s5}, randomly generated small scales are also added to the spectral parameter templates.

From Fig.~\ref{fig:P_fgs_vs_freq} we can see the deviations of \texttt{d1} (\texttt{s1}) and \texttt{d10} (\texttt{s5}) per pixel SEDs from the spatially constant ones of \texttt{d0} (\texttt{s0}).
The models \texttt{d1}, \texttt{s1} and \texttt{d10}, \texttt{s5} are indeed the ones motivating the development of component separation techniques robust against the spatial variations of the foreground SEDs, as they present different levels of spatial variability of their spectral parameters and are currently both considered viable representations of the complexity of the thermal dust and synchrotron emissions~\citep{borrill2025full}.
These are the models we study in the following.

\begin{figure*}
    \centering
    \includegraphics[width=\textwidth]{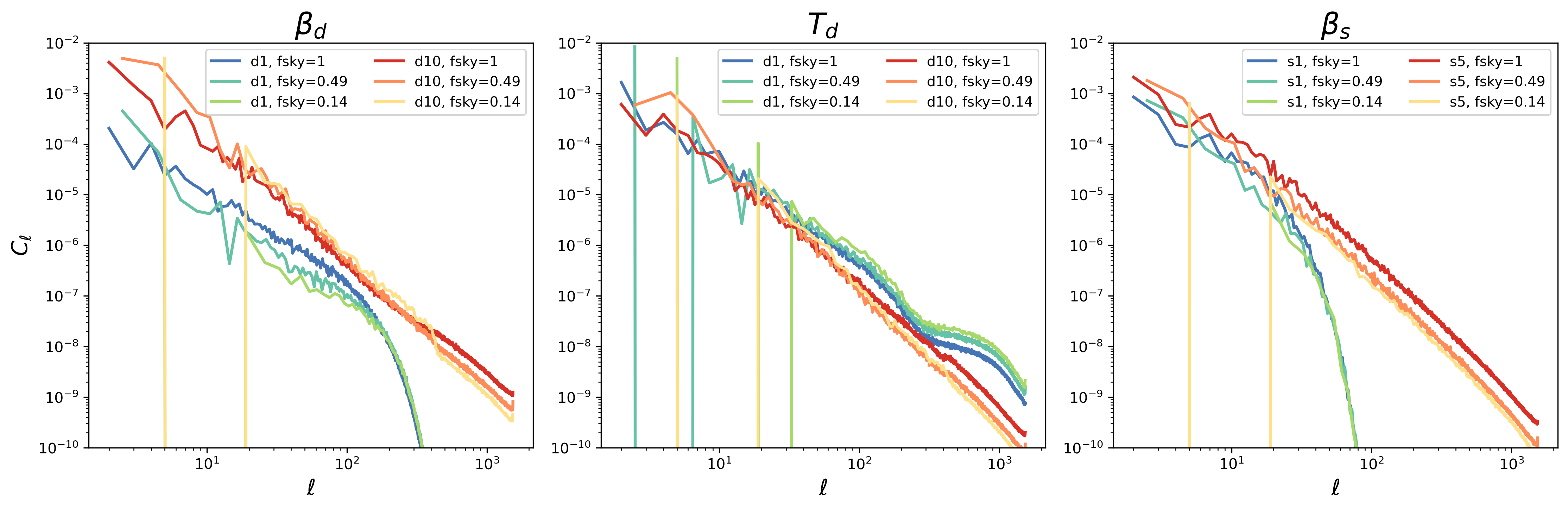}
    \caption{Angular power spectra of the three spectral indices studied in this work, \Bd, \Td\ and \Bs, computed for various sky cuts (progressively masking more and more the Galactic plane), both for the \texttt{d1} (\texttt{s1}) and \texttt{d10} (\texttt{s5}) \texttt{PySM3} models.
    The power spectra are computed with \texttt{NaMaster}~\citep{alonso2019unified}, and by apodizing the Planck HFI Galactic plane masks~\citep{HFI_Mask} with the ``Smooth'' apodization of \texttt{NaMaster} and an apodization scale of $2.5^\circ$. A binning of $\Delta\ell=2$ is applied for the $\mathrm{fsky}=0.49$ mask and of $\Delta\ell=7$ for the $\mathrm{fsky}=0.14$ mask. 
    Note that, before computing the power spectrum of the \Td\ maps, we subtract the mean of each map and then divide by it, in order to bring the resulting power spectrum into the same range of values as those of \Bd\ and \Bs, and to make it dimensionless.}
    \label{fig:Cl_beta}
\end{figure*}

We show in Fig.~\ref{fig:Cl_beta} the angular power spectra of the $\Bs$, $\Bd$ and $\Td$ templates, both for \texttt{d1} (\texttt{s1}) and \texttt{d10} (\texttt{s5}), for three sky areas, with growing fraction of the sky masked along the Galactic plane: a full-sky case, a $f_{\rm sky}=49\%$ mask and a $f_{\rm sky}=14\%$ mask\footnote{Note that the masks used here are the Planck HFI Galactic plane masks~\citep{HFI_Mask} with $f_{\rm sky}=20\%$ and $f_{\rm sky}=60\%$ which, for the power spectrum computation, have been apodized with the ``Smooth'' apodization of \texttt{NaMaster}~\citep{alonso2019unified} and apodization scale of $2.5^\circ$.}.
The power-spectrum of spectral indices is rapidly decreasing with $\ell$.
This motivates the component separation techniques to characterize the spatial variations of the SEDs at least on the largest scales in order to avoid significant mistakes in their modeling and consequently large foreground residuals, as we will also see in the following sections.
In addition, we notice that, for some of the spectral indices, larger fractions of the sky correspond to higher amplitude of the power spectrum (corresponding to larger fluctuations in the template of spectral indices), for example for $\Bd$ in the \texttt{d1} model.
However, this is not always the case, for example for $\Td$ in the \texttt{d1} model, indicating that these foreground models encode spatial variability of the scaling laws extending to high Galactic latitudes.

\subsection{Systematic and statistical errors in parametric component separation: a toy model}
\label{sec:feeling}
To illustrate how the component separation works, let us assume an experiment with two-frequency channels, observing a CMB+dust+noise only sky, at $150$ and $500$~GHz.
Assuming that the contribution of the foreground components to a given frequency map is well described by a template rescaled by a frequency-dependent coefficient, the observed frequency maps can be expressed in CMB units as:
\noindent
\begin{align}
        d_{150} &= s^{\rm CMB} + a_{150}s^{\rm dust} + n_{150}\nonumber\\
        d_{500} &= s^{\rm CMB} + a_{500}s^{\rm dust} + n_{500}
        \label{eq:ex_data_model}
\end{align}
with $s^{\rm CMB}$ the CMB map amplitude, $a_\nu s^{\rm dust}$ the dust map amplitude at a given frequency $\nu$ and $n_\nu$ the noise map at this frequency. 
Provided an estimate of the dust frequency-scaling law, $\Bar{a}_\nu$, the component separation builds an estimate of the CMB map by rejecting the dust contamination:
\begin{equation}
        \Bar{s}^{\rm CMB} = \frac{d_{150} - \Bar{a}_{150}d_{500}/\Bar{a}_{500}}{1 - \Bar{a}_{150}/\Bar{a}_{500}}.
        \label{eq:easy_compsep}
\end{equation}
The difference between the recovered CMB signal, $\Bar{s}^{\rm CMB}$, and the true CMB sky, $s^{\rm CMB}$, corresponds to the residuals and reads
\begin{align}
\delta s^{\rm CMB} &\equiv \Bar{s}^{\rm CMB} - s^{\rm  CMB} \\
&\equiv \delta s^{\rm dust} + {\rm noise\ term}.
\label{eq:example_res_decomposition}
\end{align}
In this simple example we can write analytical expressions for both the noise term and the dust residuals by combining Eqs.~(3), (4) and (5).
We find that the noise term is given by
\begin{equation}
    {\rm noise\ term} =
    \frac{\Bar{a}_{500} n_{150} - \Bar{a}_{150}n_{500}}{\Bar{a}_{500} - \Bar{a}_{150}},
\end{equation}
and the dust residuals, $\delta s^{\rm dust}$, are:
\begin{equation}
    \delta s^{\rm dust} = \left( a_{150} - \Bar{a}_{150} \frac{a_{500} - a_{150}}{\Bar{a}_{500} - \Bar{a}_{150}} \right) s^{\rm dust}.
    \label{eq:delta_s_dust}
\end{equation}
We see that the noise maps coming from both frequency channels contribute to the noise after component separation and, being uncorrelated, lead to a boosted noise variance in the CMB reconstructed map.
On the other hand, the dust residual term vanishes when the dust frequency scaling is correctly estimated. There are two possible sources of mismatch between $\Bar{a}_{\nu}$ and $a_{\nu}$: a statistical error due to the noise, and a systematic error arising when the parametric scaling laws and/or their degrees of freedom are different in the component separation and the actual foregrounds.

\subsection{Single-pixel parametric maximum-likelihood component separation}

Consider a multi-frequency observation for a given sky pixel. Analogously to Eq.~\eqref{eq:ex_data_model}, we collect all the measurements into a frequency vector $\mathbf{d}$ that we model as a linear combination of the unknown CMB and foregrounds --- which we stack in a component vector $\mathbf{s}$ --- and a Gaussian distributed white noise $\mathbf{n}$ with covariance $\mathbf{N}$:
\begin{align}
    \mathbf{d} = \mathbf{A}(\bm{\beta}) \, \mathbf{s} + \mathbf{n}.
    \label{eq:data_model}
\end{align}
The mixing matrix $\mathbf{A}$ collects in its columns the models for the SED of the components (i.e. the $a_{\nu}$ of the previous section), which depend on a set of free foreground parameters that we collect in the vector $\bm{\beta}$.
The spectral parameters $\bm{\beta}$ and component maps $\mathbf{s}$ can be estimated by maximizing the likelihood
\begin{align}
    - 2 \ln \mathcal{L}(\mathbf{s}, \bm{\beta}) = \left[\mathbf{d} - \mathbf{A}(\bm{\beta}) \, \mathbf{s} \right]^\top \mathbf{N}^{-1} \left[\mathbf{d} - \mathbf{A}(\bm{\beta}) \, \mathbf{s} \right] + {\rm const}.
    \label{eq_data_likelihood}
\end{align}

The Fisher matrix for the $\{ \mathbf{s},\,\bm{\beta} \}$ parameters belonging to pixel $p$ is given by
\begin{align}
    \label{eq:fisher_general_def}
    \mathbf{I} &\equiv \left. \left\langle \frac{\partial^2 \mathrm{ln}\, \mathcal{L}}{\partial x \partial y}\right\rangle \right|_{\substack{\{x,y\}\in\{\mathbf{s},\bm{\beta}\}\\\mathbf{s}=\mathbf{s}^{\rm true},\,\bm{\beta}=\bm{\beta}^{\rm true}}}\\
    &= \left[ \begin{array}{c|c|c}
     & \\
    \mathbf{A_{c}} & \mathbf{A_{f}} & \mathbf{D} \\
     & 
    \end{array} \right]^{\top} \mathbf{N}^{-1} \left[ \begin{array}{c|c|c}
     & \\
    \mathbf{A_{c}} & \mathbf{A_{f}} & \mathbf{D} \\
     &
    \end{array} \right],
    \label{eq:fisher}
\end{align}
where we have split the mixing matrix $\mathbf{A}$ into two matrices, $\mathbf{A_{c}}$ containing the SED of the CMB, and $\mathbf{A_{f}}$ containing the SED of the foregrounds. 
Assuming that the CMB frequency scaling is perfectly known, which is a good approximation as its spectrum has been measured to high precision by the FIRAS experiment \cite{Fixsen:1996nj}, we can express the amplitudes of all components in CMB units. In these, the CMB amplitude is constant with frequency, i.e. $\mathbf{A_{c}}$ is just filled with ones. We have also introduced $\mathbf{D} \equiv (\partial_{\beta} \mathbf{A} ) \, \mathbf{s}$, which contains the first order derivative of the data model with respect to the entries of $\bm{\beta}$, and thus depends only on the foreground part $\mathbf{A_{f}}$ as the CMB part is constant.
When we compute $\mathbf{I}$, the mixing matrix and its derivatives are estimated at the true values of $\bm{\beta}$, which is indeed what makes all the terms with $\left \langle \mathbf{d} - \mathbf{A} \mathbf{s}\right \rangle$ to vanish when going from Eq.~\eqref{eq:fisher_general_def} to Eq.~\eqref{eq:fisher}.

From Eq.~\eqref{eq:fisher} we can estimate the CMB post-component separation noise power to be
\begin{equation}
    \sigma^{2} =  \left[\mathbf{I}^{-1}\right]_{{\rm CMB}, {\rm CMB}} = \left( \mathbf{A_{c}}^{\top} \mathbf{P_{f}} \mathbf{A_{c}} \right)^{-1},
    \label{eq:noise}
\end{equation}
where $\mathbf{P_{f}}$ is the projector orthogonal to $\mathbf{B_{f}} \equiv \left[ \mathbf{A_{f}} | \mathbf{D} \right]$ defined as
\begin{equation}
    \mathbf{P_{f}} \equiv \mathbf{N}^{-1} - \mathbf{N}^{-1}\mathbf{B_{f}} \left( \mathbf{B_{f}}^\top\mathbf{N}^{-1}\mathbf{B_{f}} \right)^{-1} \mathbf{B_{f}}^\top\mathbf{N}^{-1}.
    \label{eq:projector}
\end{equation}
Note that $\mathbf{P_{f}}$ does not depend on the normalization of the columns of $\mathbf{B_f}$, but only on the space they span. 
In particular, this means that the amplitude of the foregrounds $\mathbf{s}$ is irrelevant because it only affects the normalization of the columns of $\mathbf{D}$. The statistical uncertainty associated to the recovered CMB map is therefore independent on the amplitude of Galactic foregrounds.

$\mathbf{I}$ is the information provided by the multi-frequency measurements. 
The total information on the CMB is the contribution to the inverse of Eq.~\eqref{eq:noise} coming from the first term of Eq.~\eqref{eq:projector} ($\mathbf{N}^{-1}$).
While the second term of Eq.~\eqref{eq:projector} gives, in Eq.~\eqref{eq:noise}, the portion that is degenerate with the other parameters we are fitting for (foreground amplitudes and spectral parameters) and inflates the statistical uncertainty in the CMB estimates.
The role of this second term in $\mathbf{P_{f}}$ is indeed to set $\mathbf{I}$ (the total information) to zero on a number of dimensions equal to the linear and non-linear parameters.
One may be tempted to think that, if one more foreground parameter is added to the model, the consequent information loss is 1 over the number of frequencies. 
However, the reality is typically worse. The 
reason being that the CMB, the foreground SEDs and their derivatives are typically smooth and, therefore, few parameters are sufficient to make them degenerate.
Another perspective on the same problem is that the statistical gain obtained by introducing many frequencies is limited if the foreground SEDs and their derivatives are smooth.

To summarize, when using the likelihood in Eq.~\eqref{eq_data_likelihood}, 1) the information loss due to the component separation only 
depends on how orthogonal the CMB SED $\mathbf{A_c}$ is to the foreground SEDs $\mathbf{A_f}$ and their derivatives $\mathbf{D}$ with respect to the spectral parameters: it does not depend on the amplitude of the foregrounds; 2) every addition of a free spectral parameter degrades the post-component separation sensitivity, often substantially. These remarks remain valid for both parametric and blind approaches, provided they use the likelihood of Eq.~\eqref{eq_data_likelihood}. The parameters in the second case would be, in practice, the elements of $\mathbf{A_f}$, however a prior would typically be added to Eq.~\eqref{eq_data_likelihood}, which would improve the separability of the CMB and foregrounds (e.g.~\cite{Leloup:2023vkb}).

\section{Spectral likelihood based component separation}
\label{sec:sp_lik_based_compsep}
We now move on to describe the particular case of the \fgb\ component separation.
We first summarize the component separation core steps, and then present a semi-analytical framework to efficiently estimate the CMB residuals independently of noise realizations.
These discussions build on \cite{Stompor:2008sf} and \cite{Stompor:2016hhw}, extending these works to pixel dependent mixing matrices as implemented in \fgb.
Also, we especially discuss the nature and dependence of the post component separation residuals, which we then validate in Sect.~\ref{sec:illustration}.

\subsection{\fgb: simulation based}
\label{sec:sim_based_framework}
In a full-sky observation, for each pixel we have a multi-frequency observation $\mathbf{d}_p$ and a vector of component (CMB and foregrounds) amplitudes $\mathbf{s}_p$.
Although the mixing matrix is applied in each pixel $p$, the spectral parameters themselves are defined on generic pixel subsets. 
In other words, we assume to have the same value $\beta_i(\mathcal{P}_i)$ in all pixels of a generic pixel subset for each spectral parameter type $i$, typically representing \Bd, \Td, \Bs.
In the following we write this concisely as $\bm{\beta}_{\bm{\mathcal{P}} \left( p \right)}$, highlighting the dependence of $\bm{\mathcal{P}}$ on the sky pixel considered $p$. 
The likelihood then becomes
\begin{align}
    &- 2 \ln \mathcal{L}(\{\mathbf{s}_p\}, \{\bm{\beta}_{\bm{\mathcal{P}} \left( p \right)}\}) = \nonumber\\ &\hspace{.5cm}\sum_p \left[\mathbf{d}_p - \mathbf{A}(\bm{\beta}_{\bm{\mathcal{P}} \left( p \right)}) \mathbf{s}_p \right]^\top \mathbf{N}^{-1} \left[\mathbf{d}_p - \mathbf{A}(\bm{\beta}_{\bm{\mathcal{P}} \left( p \right)}) \mathbf{s}_p \right] + {\rm const}.
    \label{eq:like_pixel_sum}
\end{align}
This expression assumes that the noise is homogeneous across the sky, and uncorrelated between pixels. Although computationally more intensive, this equation can in principle be implemented for any Gaussian noise behaviours.
To minimize Eq.~\eqref{eq:like_pixel_sum} (i.e. maximize the likelihood $\mathcal{L}$) the fit can happen independently for pixels not sharing the pixel subset for any of the spectral parameter types, but must happen jointly for pixels sharing at least one spectral parameter type.
We note that the mixing matrix is different as soon as a single spectral parameter changes, so $\mathbf{A}$ can vary over smaller patches than the $\mathcal{P}_i$ if these are distinct between parameter types (here \Bd, \Td or \Bs). 
For clarity to the expense of precision, we only express the dependence of the mixing matrix with the pixel $p$ in the following. Because of the pixel dependence of the mixing matrix, the noise of the reconstructed CMB is not homogeneous anymore, but is still uncorrelated between pixels. In each pixel, it is given by Eq.~\eqref{eq:noise}.

As previously demonstrated, the performance of the component separation does not depend on the amplitude of the signal, so we get rid of the parameters $\left\{ \mathbf{s}_{p} \right\}$. 
This is achieved by maximizing the likelihood with respect to $\mathbf{s}_{p}$ to obtain the spectral likelihood:
\begin{align}
    &- 2 \ln \mathcal{L}(\{\bm{\beta}_{\bm{\mathcal{P}} \left( p \right)}\}) = \nonumber\\ &\hspace{.5cm}-\sum_p \mathbf{d}_p^\top \mathbf{N}_{p}^{-1} \mathbf{A}_p \left( \mathbf{A}_p^\top \mathbf{N}_{p}^{-1} \mathbf{A}_p \right)^{-1} \mathbf{A}_p^\top \mathbf{N}_{p}^{-1} \mathbf{d}_p + {\rm const}.
    \label{eq:spec_like_pixel_sum}
\end{align}
We maximize instead of marginalizing because it was shown in~\citep{Stompor:2008sf} that the spectral likelihood remains unbiased in the spectral parameters. The maximum likelihood signal component $\mathbf{s}_{p}$ is given by the generalized least-square estimate in this context:
\begin{align}
    \mathbf{\Bar{s}}_{p} = \left( \mathbf{\Bar{A}}_{p}^\top\ \mathbf{N}^{-1} \mathbf{\Bar{A}}_{p} \right)^{-1} \mathbf{\Bar{A}}_{p}^\top\ \mathbf{N}^{-1} \mathbf{\hat{d}}_{p},
\label{eq:gls}
\end{align}
where the best-fit parameters $\bm{{\Bar{\beta}}}$ determine the best-fit mixing matrix $\mathbf{\Bar{A}}_p$, which we then use for the estimation of the maximum-likelihood component amplitudes.
The implementation of these two steps has been released already in ~\citep{2023ascl.soft07021P}, hence we just give its description, together with some technical aspects of the likelihood maximization, in  appendix~\ref{ap:implementation}.
In the following instead we focus on the interpretation of the component separation residuals.

Fitting for an extra free parameter comes at a considerable cost and should be avoided whenever possible. 
That said, additional spectral parameters are often necessary to reduce the bias on cosmological parameters induced by the presence of foreground residuals~\citep[see e.g.][]{errard2019characterizing}. 
We can extend the discussion introduced in Sect.~\ref{sec:feeling} for the hypothetical two-frequency-channel experiment to this more general description, replacing Eq.~\eqref{eq:data_model} in Eq.~\eqref{eq:gls} and focusing on the CMB component:
\begin{align} 
    \delta \mathbf{s}_p^{\mathrm{CMB}} &\equiv \mathbf{\Bar{s}}_p^{\mathrm{CMB}} - \mathbf{s}_p^{\mathrm{CMB}} \label{eq:res definition} \\
    &= \left[ (\mathbf{\Bar{A}}_p^\top\ \mathbf{N}^{-1} \mathbf{\Bar{A}}_p)^{-1} \mathbf{\Bar{A}}_p^\top\ \mathbf{N}^{-1} \mathbf{A}\left(\bm{\beta}_{p}\right) \right]_\mathrm{fg} \mathbf{s}_p^\mathrm{fg} + \mathrm{noise \, term}, \label{eq:res decomposition}
\end{align}
hence generalizing Eq.~\eqref{eq:example_res_decomposition} to:
\begin{equation}
    \delta \mathbf{s}^{\mathrm{CMB}}_p \equiv \delta \mathbf{s}^{\mathrm{fg}}_p + \mathrm{noise \, term}.
\end{equation}
To obtain Eq.~\eqref{eq:res decomposition} from Eq.~\eqref{eq:res definition}, we assume that the CMB frequency scaling is perfectly known.

Again, the foreground residual term vanishes if the true mixing matrix $\mathbf{A(\bm{\beta})}$ is correctly estimated, and, in a given pixel, there are two possible sources of mismatch between the true spectral parameters ${\bm \beta}_{p}$ and the estimated best-fit ones $\bm{\Bar{\beta}}_{\bm{\mathcal{P}} \left( p \right)}$: a statistical error due to the noise, and a systematic error arising if the parametric scaling laws and/or their degrees of freedom are different in the component separation and the actual foregrounds.
In other words, we distinguish between statistical and systematic errors on the estimated parameters of the scaling laws, 
\begin{equation}
\delta \bm{\beta}=\delta \bm{\beta}_{\rm stat}+\delta \bm{\beta}_{\rm syst}.
\end{equation}
We can understand intuitively that this decomposition must exist, because of the independence of the two effects mentioned, however we also verify this on simulations in Sect.~\ref{subsect:stat_syst_errors_on_betas}. 
In the limit where one of the effects vanishes, the other still exists and leads to errors.
We observe that, although $\delta \bm{\beta}$ varies from pixel to pixel, the dependence of the statistical part (on each pixel subset) and of the systematic part (on each pixel) is distinct due to their different origins.

Because Eq.~\eqref{eq:gls} is non-linear in the spectral parameters, it is not possible to analytically decompose $\delta \mathbf{s}^{\rm fg}$ into two terms, each depending only on $\delta \bm \beta_{\rm syst}$ or $\delta \bm \beta_{\rm stat}$.
However, we can still distinguish between statistical and systematic residuals at the map level, 
\begin{equation}
\delta \mathbf{s} = \delta \mathbf{s}_{\mathrm{syst}}(\bm \beta_{\mathrm{syst}})+ \delta \mathbf{s}_{\mathrm{\mathrm{stat}}}(\bm \beta_{\mathrm{syst}}, \bm \beta_{\mathrm{stat}}),
\end{equation}
where delta $\delta \mathbf{s}_{\mathrm{syst}}$ is the remaining residuals in the limit of vanishing noise, which in practice is obtained as the residual map after averaging over different noise realizations. On the other hand, delta $\delta \mathbf{s}_{\mathrm{stat}}$ is the noise realization-specific part of the residual map.
The former term, when the average is done on component separation results from a large number of different noise realizations, is indeed independent from $\delta \bm \beta_{\rm stat}$ and hence does depend on $\delta \bm \beta_{\rm syst}$ only.
This can be intuitively understood, but we also show it in Sect.~\ref{subsubsection:Systematic residuals}.
The term $\delta \mathbf{s}_{\mathrm{stat}}$ instead generally depends both on $\delta \bm \beta_{\rm stat}$ and $\delta \bm \beta_{\rm syst}$. Due to a common misconception the latter dependence is frequently overlooked.

Relying on an average over the result from performing the component separation over different noise realizations is only possible when working with simulations and it would not be directly transposable in a real data analysis scenario. 
In the latter case we cannot access or estimate the systematic foreground residual contribution.
We need instead an estimate for the statistical contribution to the residuals.
For the latter we could use a similar approach, based on direct sampling of the likelihood, that would allow us to access the statistical residuals even if at a higher computational cost.
Characterizing the spectral likelihood through sampling, instead of accessing its maximum only, would give us a chain of spectral parameter estimates, using those we would hence get a set of recovered CMB maps, containing both the true CMB, systematic and statistical residuals.
From those and with the same procedure described above we could then extract an estimate of the statistical contribution to the residuals.

\subsection{\fgb: forecasting framework}
\label{sec:forecasting}

\subsubsection{Motivation}

Aside from performing the component separation on actual observed data, another crucial need in the early stages of an instrument development is the production of performance forecasts.
Such a task requires tools that can perform fast but reliable cosmological parameter estimates starting from specific instrumental characteristics.
Data products such as component separated maps are not needed, so some steps may be streamlined to accelerate the procedure. 
It is therefore common to adapt the general framework to the particular needs of a forecasting approach \citep[see e.g.][]{Stompor:2016hhw,Leloup:2025kaw}, and this is the focus of the present section.

This forecasting framework follows a semi-analytical, Fisher-like formalism where the performance of the component separation is estimated on average in a quick and efficient way without the need to produce potentially expensive noise simulations. 
It will therefore be a suitable tool when exploring the characteristics of a large number of different instrument configurations, or of the hyper-parameters of the component separation scheme itself, such as the number of patches and their morphology, the masking procedure, etc.

Finally, to develop such an approach, it is necessary to model the method’s response and break down the problem into its elementary blocks, offering more insight into its important aspects and those that are irrelevant than running over simulations.
For example, regarding the topic discussed in this work, this framework can provide useful information on the spatial properties, distribution and correlations, of the residual foregrounds as a function of the specifics of the experiment.

\subsubsection{Estimation of residuals}
The formalism builds on~\cite{Stompor:2016hhw}. 
After briefly summarizing the latter work, we describe how we adapt it to account for possibly pixel-dependent spectral properties of the foregrounds. 
We start from the spectral likelihood in Eq.~\eqref{eq:spec_like_pixel_sum} which, after averaging over noise realizations and a few simplifications, can be expressed as:
\begin{align}
    &-2 \left< \ln \mathcal{L}_{\rm spec} \right> = \nonumber\\ 
    &\hspace{.5cm}- \sum_p \mathbf{\hat{d}}_p^\top\ \mathbf{N}^{-1} \mathbf{A}_p (\mathbf{A}_p^\top\ \mathbf{N}^{-1} \mathbf{A}_p)^{-1} \mathbf{A}_p^\top\ \mathbf{N}^{-1} \mathbf{\hat{d}}_p + {\rm const},
    \label{eq:avg_spec_like}
\end{align}
where the hat corresponds to taking a noiseless realization of the observed data. In other words, the averaged likelihood is simply the likelihood itself, applied to noiseless observations which can be taken to be foregrounds only since including the CMB component has no effect on the residuals, in the absence of instrumental systematics (in cases where systematics are present, see e.g.~\cite{verges2021framework,jost2023characterizing,leloup2024impact}).

The reconstructed component maps from the averaged likelihood $\mathbf{\Bar{s}}$ are given by the Generalized Least-Square estimate of Eq.~\eqref{eq:gls} evaluated using the best-fit parameters of the averaged likelihood $\bm{\Bar{\beta}}_{\bm{\mathcal{P}}(p)}$.
Introducing the weighting operator $\mathbf{W}_{p}$ that has the same spatial dependence as the mixing matrix $\mathbf{A}$, we can rewrite the reconstructed maps as
\begin{equation}
    \mathbf{\Bar{s}}_{p} = \mathbf{W}_{p} \left( \bm{\Bar{\beta}}_{\bm{\mathcal{P}} \left( p \right)} \right)\, \mathbf{\hat{d}}_{p}.
\end{equation}
Since the input maps only contain foregrounds here, what is reconstructed as CMB by the component separation is by definition the noiseless residuals, i.e. the systematic residuals $\delta \mathbf{s}_{\mathrm{syst} \, p}$.
Therefore, these residuals are simply the CMB component of the reconstructed maps:
\begin{equation}
    \delta \mathbf{s}_{\mathrm{syst} \, p} = \left[ \mathbf{\Bar{s}}_{p} \right]_{\rm CMB} = \mathbf{W}_{p}^{\rm CMB} \left( \bm{\Bar{\beta}}_{\bm{\mathcal{P}} \left( p \right)} \right)\, \mathbf{\hat{d}}_{p},
\end{equation}

The core idea of the approach is to assume that the statistical properties of these residuals, as noise realizations vary, can be captured by Taylor expanding the weighting operator in the spectral parameters around their maximum averaged likelihood values $\{ \bm{\Bar{\beta}} \}$. In other words, the presence of noise is expected to introduce small variations $\delta \bm{\beta}$ in the spectral parameters around the recovered ones in the noiseless case. Up to second order in the perturbation of the spectral parameters, we can thus write the residuals as:
\begin{equation}
    \delta \mathbf{s}_{p} = \left( \mathbf{W}_{p}^{\left( 0 \right)} + \mathbf{W}_{p, \beta}^{\left( 1 \right)} \delta \bm{\beta}_{\bm{\mathcal{P}} \left( p \right)} + \delta \bm{\beta}_{\bm{\mathcal{P}} \left( p \right)}^\top\ \mathbf{W}_{p, \beta\beta'}^{\left( 2 \right)} \delta \bm{\beta}'_{\bm{\mathcal{P}} \left( p \right)} \right) \mathbf{\hat{d}}_{p} \label{eq:residual_map}.
\end{equation}

In the above expression, we introduced the 0th, 1st and 2nd order operators of the expansion of the weighting operator respectively defined as:
\begin{align}
    &\mathbf{W}_{p}^{\left( 0 \right)} \equiv \mathbf{W}_{p}^{\rm CMB} \left( \Bar{\beta}_{\bm{\mathcal{P}} \left( p \right)} \right) 
    \label{eq:W_0}\\
    &\mathbf{W}_{p, \beta}^{\left( 1 \right)} \equiv \left. \frac{\partial \mathbf{W}_{p}^{\rm CMB}}{\partial \beta_{\bm{\mathcal{P}} \left( p \right)}} \right|_{\Bar{\beta}_{\bm{\mathcal{P}} \left( p \right)}} 
    \label{eq:W_1}\\
    &\mathbf{W}_{p, \beta\beta'}^{\left( 2 \right)} \equiv \left. \frac{\partial^{2} \mathbf{W}_{p}^{\rm CMB}}{\partial \beta'_{\bm{\mathcal{P}} \left( p \right)} \partial \beta_{\bm{\mathcal{P}} \left( p \right)}} \right|_{\Bar{\beta}_{\bm{\mathcal{P}} \left( p \right)}}.
    \label{eq:W_2}
\end{align}

If the perturbations $\delta \bm{\beta}$ are to be representative of the statistical variations in the spectral parameters, their distributions need to verify both properties:
\begin{equation}
    \left< \delta \beta_{\bm{\mathcal{P}} \left( p \right)} \delta \beta_{\bm{\mathcal{P}}' \left( p' \right)}^\top\ \right> = \bm{\Sigma}_{\bm{\mathcal{P}} \bm{\mathcal{P}}'} \qquad \text{and} \qquad \left< \delta \beta_{\bm{\mathcal{P}} \left( p \right)} \right> = 0,
    \label{eq:stat_prop_delta_beta}
\end{equation}
with $\bm{\Sigma}$ the inverse of the Hessian of the averaged pixel-dependent likelihood evaluated at its maximum, referred to in the following as the $\delta \bm{\beta}$ covariance matrix, which also corresponds to the inverse of the Fisher-like matrix defined in Eq.~\eqref{eq:xForecast Fisher} below.

From the expression of the residuals, Eq.~\eqref{eq:residual_map}, and the statistical properties of $\delta \bm{\beta}$, Eq.~\eqref{eq:stat_prop_delta_beta}, we have the analytical expression for the pixel-pixel covariance of the residuals up to second order in $\delta \bm{\beta}$:
\begin{align}
    &\mathbf{C}_{pp'} = \mathbf{\hat{d}}_{p'}^\top\ \left( \mathbf{W}_{p'}^{\left( 0 \right) \top} \mathbf{W}_{p}^{\left( 0 \right)} + \mathrm{Tr} \left[ \left( \mathbf{W}_{p'}^{\left( 1 \right)} \mathbf{W}_{p}^{\left( 1 \right)} \bm{\Sigma}_{\bm{\mathcal{PP'}}} \right)^\top\ \right] \right. \nonumber \\
    & +\, \mathbf{W}_{p'}^{\left( 0 \right) \top} \mathrm{Tr} \left[ \mathbf{W}_{p}^{\left( 2 \right)} \bm{\Sigma}_{\bm{\mathcal{PP}}} \right] + \left( \mathrm{Tr} \left[ \mathbf{W}_{p'}^{\left( 2 \right)} \bm{\Sigma}_{\bm{\mathcal{P'P'}}} \right] \right)^\top\ \mathbf{W}_{p}^{\left( 0 \right)} ) \mathbf{\hat{d}}_{p},
    \label{eq:cls_semi_analytical}
\end{align}
where the transposition acts on the dimension of $\mathbf{W}^{0}$ (frequencies) and the trace acts on the dimension of $\delta \bm{\beta}$ (types of spectral parameters).

Although this pixel-pixel covariance matrix \textit{a priori} contains valuable information for understanding the statistical properties of the residuals and their spatial correlations, especially in the general case of pixel dependent component separation, its terms can be difficult to compute in practice due to the potentially large size of the involved quantities.
In addition, it is unclear how this covariance is related to statistical properties of the angular distribution on the sphere, in particular the angular power spectrum of residuals.

Alternatively, instead of pursuing an analytical description of the statistical properties of residuals, it is possible to get a fast and accurate estimate of their angular power spectrum if we accept a semi-analytical approach. 
Each of the terms in Eq.~\eqref{eq:residual_map} can then be computed from random realizations of $\delta \bm{\beta}$, which can be generated assuming they verify the properties of Eq.~\eqref{eq:stat_prop_delta_beta}. 
This approach corresponds to producing realizations of fake residuals from sets of $\delta \bm{\beta}$ randomly generated from their covariance $\bm{\Sigma}$.
The statistical properties of these fake residuals then follow those of the actual residuals. This is the approach that we describe in the following.

\subsubsection{Implementation}

There are two main steps to describe for the forecasting approach:
\begin{enumerate}
    \item the calculation of the Fisher-like matrix $\bm{\mathcal{I}}$ from the averaged pixel-dependent likelihood of Eq.~\eqref{eq:avg_spec_like},
    \item the generation of realizations of $\delta \bm{\beta}$ following the properties in Eq.~\eqref{eq:stat_prop_delta_beta}.
\end{enumerate}

The Fisher-like matrix is defined from the second derivatives of the averaged likelihood:
\begin{equation}
    \bm{\mathcal{I}} = \bm{\Sigma}^{-1} = \left. \left< \frac{\partial^{2} \ln \mathcal{L}_{\rm spec}}{\partial \beta'_{\mathcal{P'} \left( p' \right)} \partial \beta_{\mathcal{P} \left( p \right)}} \right> \right|_{\Bar{\beta}_{\mathcal{P} \left( p \right)}} \label{eq:xForecast Fisher}
\end{equation}

A derivation of the Hessian of the average spectral likelihood can be found in \citep{Stompor:2016hhw} with a missing factor of $-2$ in the result. We give here the correct expression for completeness:
\begin{align}
    & \left< \frac{\partial^{2} \ln \mathcal{L}_{\rm spec}}{\partial \beta'_{\mathcal{P'} \left( p' \right)} \partial \beta_{\mathcal{P} \left( p \right)}} \right> = \nonumber \\
    & \quad \sum_{p} \mathrm{Tr} \left[ \left( \mathbf{N}^{-1} \mathbf{A}_{p} \mathbf{N}^{\mathbf{A}}_{p} \mathbf{A}^\top_{p, \beta'} \mathbf{N}^{-1} \mathbf{A}_{p} \mathbf{N}^{\mathbf{A}}_{p} \mathbf{A}^\top_{p, \beta} \mathbf{P}_{p} \right. \right. \nonumber \\
    & \quad \quad \quad \quad + \mathbf{N}^{-1} \mathbf{A}_{p} \mathbf{N}^{\mathbf{A}}_{p} \mathbf{A}^\top_{p, \beta} \mathbf{N}^{-1} \mathbf{A}_{p} \mathbf{N}^{\mathbf{A}}_{p} \mathbf{A}^\top_{p, \beta'} \mathbf{P}_{p} \nonumber \\
    & \quad \quad \quad \quad - \mathbf{P}_{p} \mathbf{A}_{p, \beta'} \mathbf{N}^{\mathbf{A}}_{p} \mathbf{A}^\top_{p, \beta} \mathbf{P}_{p} - \mathbf{N}^{-1} \mathbf{A}_{p} \mathbf{N}^{\mathbf{A}}_{p} \mathbf{A}^\top_{p, \beta \beta'} \mathbf{P}_{p} \nonumber \\
    & \quad \quad \quad \quad \left. \left. + \mathbf{N}^{-1} \mathbf{A}_{p} \mathbf{N}^{\mathbf{A}}_{p} \mathbf{A}^\top_{p, \beta'} \mathbf{P}_{p} \mathbf{A}_{p, \beta'} \mathbf{N}^{\mathbf{A}}_{p} \mathbf{A}^\top_{p} \mathbf{N}^{-1} \right) \left< \mathbf{\hat{d}}_{p} \mathbf{\hat{d}}_{p}^\top \right> \right] \mathbf{N}^{\mathbf{A}}_{p}
\end{align}
where we have defined the projection operator $\mathbf{P}$ on the subspace orthogonal to the columns of $\mathbf{A}$ and the projected noise covariance $\mathbf{N^{A}}$ on the component space:
\begin{align}
    \label{eq:full_projector}
    \mathbf{P}_{p} & \equiv \mathbf{N}^{-1} - \mathbf{N}^{-1}\mathbf{A}_{p} \left( \mathbf{A}_{p}^\top\mathbf{N}^{-1}\mathbf{A}_{p} \right)^{-1} \mathbf{A}_{p}^\top\mathbf{N}^{-1} \\
    \mathbf{N}^{\mathbf{A}}_{p} & \equiv \left( \mathbf{A}_{p}^{\top} \mathbf{N}^{-1} \mathbf{A}_{p} \right)^{-1}
\end{align}

\begin{figure}
    \centering
    \includegraphics[width=\columnwidth]{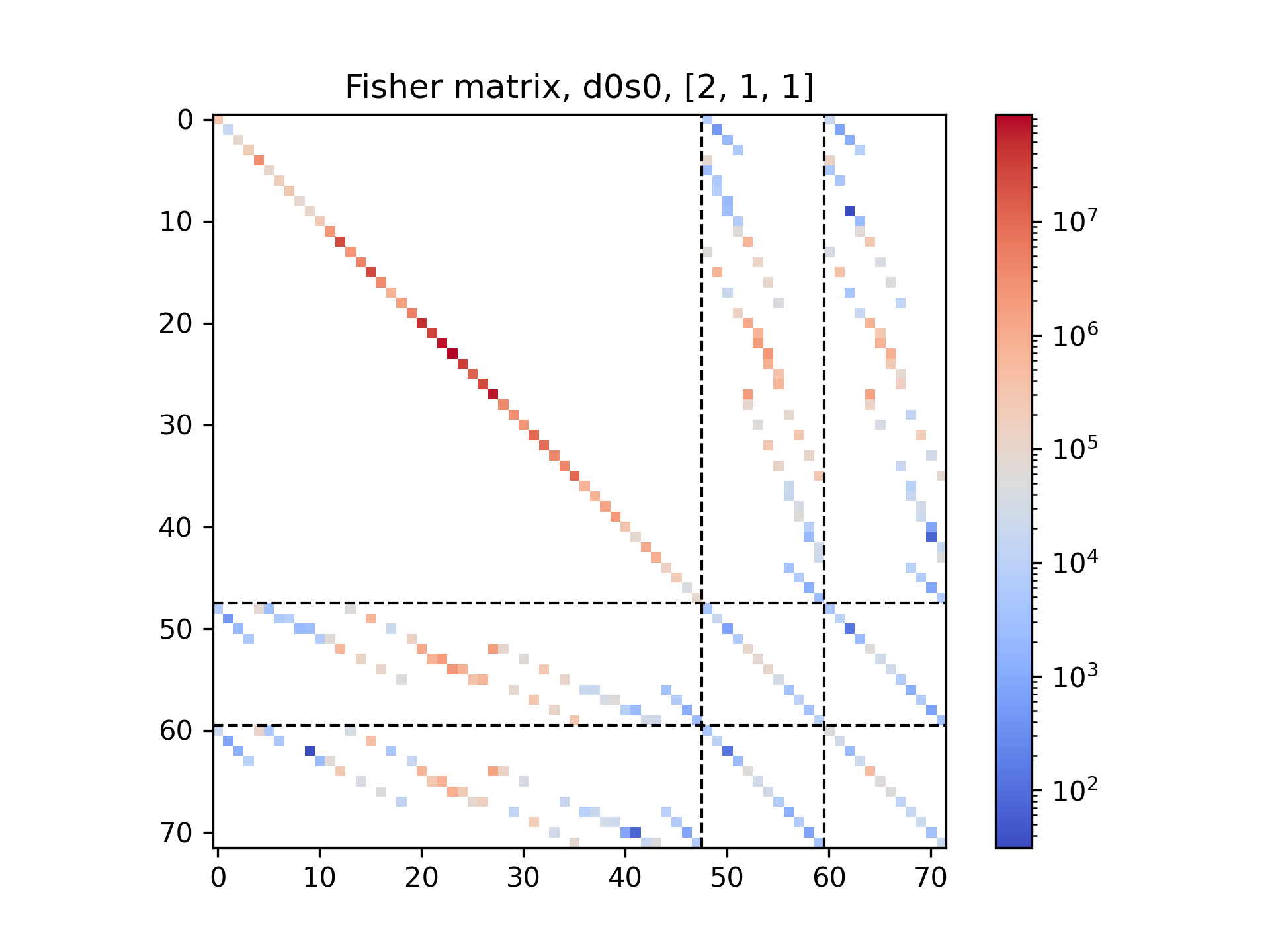}
    \caption{Fisher-like matrix for the \texttt{d0s0} foreground model full sky, with a \textit{LiteBIRD}-like setup (see~Sect.\ref{subsection:sims}), using \textsc{HEALPix} pixel patches with \texttt{nside}=[2,1,1] respectively for [$\beta_\text{d}$, $\mathrm{T}_\text{d}$, $\beta_\text{s}$]. The parameters with indices 0 to 47 correspond to $\beta_{\text{d}}$, $T_{\text{d}}$ is represented by parameters from 48 to 59 and $\beta_{\text{s}}$ by those from 60 to 71.
    Note that the entries of the Fisher-like matrix corresponding to \Bd\ and \Bs\ are dimensionless, while the ones corresponding to \Td\ are in $\rm{K}^{-2}$.}
    \label{fig:xF Fisher}
\end{figure}

Given the variability of the spectral parameters with the pixel subsets, the Fisher-like matrix is a square matrix of dimension $\sum_{i} \mathcal{P}_{i}$, with $\mathcal{P}_{i}$ the number of patches of the spectral parameter type $i = \beta_{\mathrm{d}}, \mathrm{T}_{\mathrm{d}}$ or $\beta_{\mathrm{s}}$. 
In our case, because there is no overlap between spectral parameters of the same type and because we assume that the noise covariance is diagonal in the pixel domain, the Fisher-like matrix is very sparse. 
Indeed, the diagonal blocks of size $\mathcal{P}_{i}$ are diagonal, the upper off-diagonal blocks only contain a few elements per column, corresponding to the pixel subsets of other parameter types overlapping with the pixel subset under consideration, and it is symmetric.
As an illustration of its structure, we show in Fig.~\ref{fig:xF Fisher} the Fisher-like matrix for the \texttt{d0s0} foreground model using patches defined from \textsc{HEALPix} pixels with \texttt{nside}=[2,1,1] respectively for [\Bd, \Td, \Bs].
Therefore, the Fisher-like matrix implementation makes use of the sparse linear algebra library of \texttt{SciPy}.
Once the Fisher-like matrix is obtained, we need to generate sets of random $\delta \bm{\beta}$, in such a way that they follow the two properties of Eq.~\eqref{eq:stat_prop_delta_beta}. This is achieved by finding a matrix that squares to the inverse of $\bm{\mathcal{I}}$. Several options are available and, since the Fisher-like matrix is symmetric positive definite, we settle on performing the Cholesky decomposition of $\bm{\mathcal{I}}$. The Cholesky decomposition for sparse matrices is commonly defined as finding a sparse lower triangular matrix $\mathbf{L}'$ such that $\mathbf{Q} \bm{\mathcal{I}} \mathbf{Q}^\top\ = \mathbf{L}' \left( \mathbf{L}' \right)^{\top}$ instead of the usual definition, with $\mathbf{Q}$ a permutation matrix defined to optimize the sparsity of $\mathbf{L}'$. Random sets of $\delta \bm{\beta}$ are then generated by solving the following equation:
\begin{equation}
    \left( \mathbf{L}' \right)^{\top} \mathbf{Q}\, \delta \beta = x \qquad \qquad\text{with} \quad x \sim \mathcal{N} \left( 0, 1 \right).
\end{equation}
This way, the covariance of the $\delta \beta$ is, as intended:
\begin{equation}
    \left< \delta \beta\, \delta \beta^{\top} \right> = \mathbf{Q}^{\top} \left( \mathbf{L}' \right)^{-\top} \left< x x^{\top} \right> \left( \mathbf{L}' \right)^{-1} \mathbf{Q} = \bm{\mathcal{I}}^{-1} = \bm{\Sigma}.
\end{equation}
The sparse Cholesky decomposition and solving the linear equation in $\delta \beta$ are performed using the \texttt{scikit-sparse} library.
Finally, it remains to recast the realizations of $\delta \beta$, that depend on the pixel subsets, as maps by copying the same $\delta \beta$ value in all the pixels of a given pixel subset.

For comparison with the simulation-based implementation, we define the systematic residuals as the power spectrum of the 0th order term in $\delta \bm{\beta}$ in the expansion of Eq.~\eqref{eq:residual_map}, being that term the only one independent from the statistical variations of the spectral parameters, $\delta \bm{\beta}$.
And we define the foreground statistical residuals as the power spectrum of the remaining terms (even if in practice we stop at the 1st order term only, as the 2nd order term is already negligible, see Sect.~\ref{subsection:stat_semi_analytical}), averaged over 500 realizations\footnote{We show in Sect.~\ref{subsection:stat_semi_analytical} that 500 realizations are enough to produce good agreement with the full simulation-based component separation.} of $\delta \bm{\beta}$ to which we add the power spectrum of noise after component separation. Because the mixing matrix is pixel-dependent, the noise covariance after component separation acquires a pixel dependence as it is defined by:
\begin{equation}
    \bar{N}^{\mathbf{A}} _{p} = \left[ \left( \mathbf{A}_{p}^{\top} \mathbf{N}^{-1} \mathbf{A}_{p} \right)^{-1} \right]_{\text{CMB}}.
    \label{eq:noise_cov_after_comp_sep}
\end{equation}
Once again, it is not trivial to analytically relate this covariance in the domain of pixels to the angular power spectrum of the noise after component separation. To find the latter, we therefore produce a new set of random realizations of noise after component separation as:
\begin{equation}
    n_{p} = \sqrt{\bar{N}^{\mathbf{A}} _{p}} \cdot y_{p} \qquad \qquad \text{with} \quad y \sim \mathcal{N} \left( 0, 1 \right).
\end{equation}
The angular power spectrum of noise after component separation is thus defined as the average of the power spectrum of this noise over 500 random realizations.

\section{Results: illustration of the \fgb \\component separation performance}
\label{sec:illustration}
\begin{figure}
    \centering
    \includegraphics[width=\columnwidth]{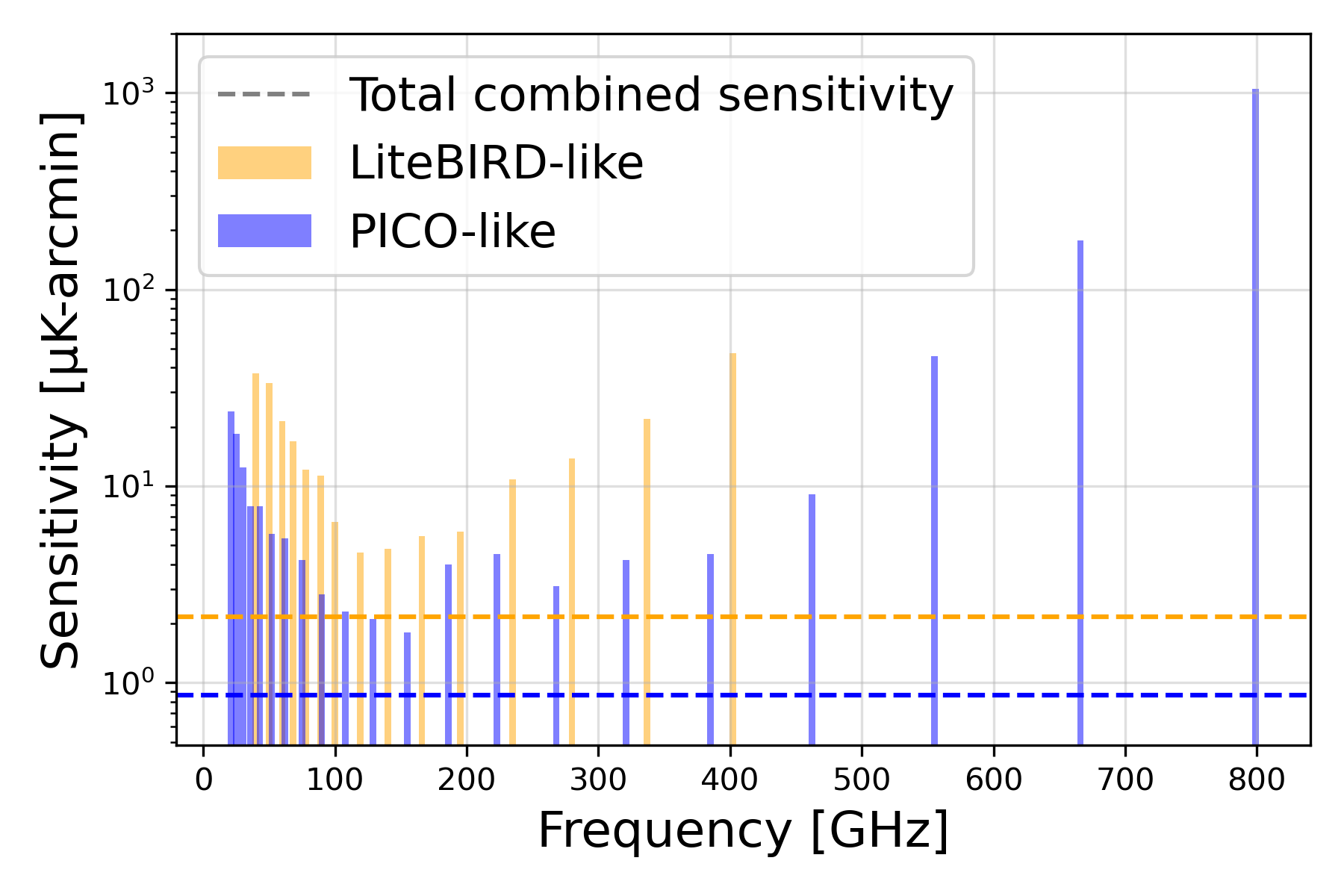}
    \caption{Frequency bands and sensitivities assumed in the analysis for the \textit{LiteBIRD}-like and \textit{PICO}-like instruments. Horizontal lines indicate the total combined sensitivity for each instrument, obtained by inverse-variance weighting of the sensitivities across the different frequency bands, yielding $2.16 \, \mu\mathrm{K\text{-}arcmin}$ and $0.86 \, \mu\mathrm{K\text{-}arcmin}$, respectively.
    Only the central frequency for each band is shown, as the bandpass widths are not relevant for the study we perform in this work.
    The key aspect relevant to the analysis and comparison of the results obtained for the two instruments is their differing frequency coverage and sensitivity levels.}
    \label{fig:LB_PICO_freq_sens}
\end{figure}

\begin{figure}
    \centering
    \includegraphics[width=\columnwidth]{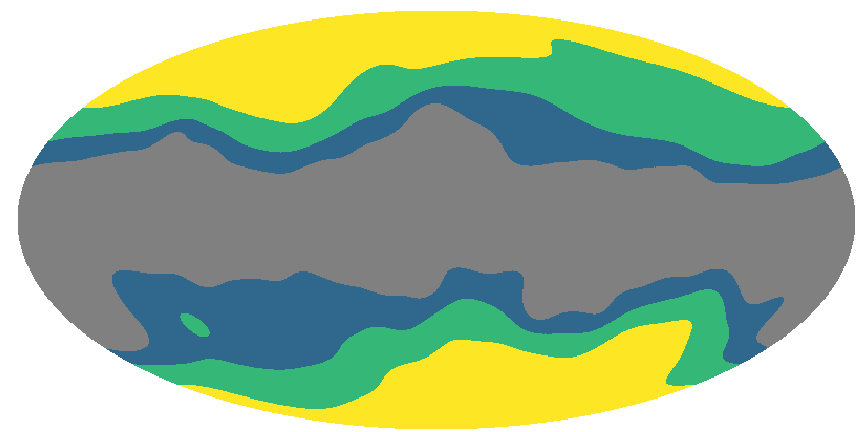}
    \caption{Illustration of the various sky masks used in this work, with fsky: $20\%$ (yellow), $40\%$ (yellow + green) and $60\%$ (yellow + green + blue).}
    \label{fig:masks}
\end{figure}

In this section we study the performance of the component separation for two space mission proposals targeting primordial $B$ modes from full-sky observations, based on the upcoming space-borne mission \textit{LiteBIRD} (LB)~\citep{litebird2023probing} and the long-term satellite proposal, \textit{PICO}~\citep{hanany2019pico}. Their specifications, such as their frequency channels and sensitivities, are given in the previous references, and we show the latter two in Fig.~\ref{fig:LB_PICO_freq_sens}. 
In a nutshell, LB will observe the whole sky in $15$ frequency bands from $40$ to $400$~GHz and target $\sigma(r=0)= 10^{-3}$; PICO plans to observe in $21$ frequency bands spread between $21$ and $799$~GHz, aiming at $\sigma(r=0) = 10^{-4}$. 
To avoid regions largely contaminated by foregrounds, our analysis in the LB-like and PICO-like scenarios is assumed to be performed across the $f_{\rm sky}=60\%$ area shown in Fig.~\ref{fig:masks}.
We focus on both scenarios as we are particularly interested in assessing how the component separation reacts to these different observational strategies: in the PICO-like observing the sky with much more sensitivity but also broader frequency range with respect to the LB-like scenario.
We expect this to give for the PICO-like case a less noisy component separation but more exposed to foreground scaling complexities.

In Sect.~\ref{subsection:sims} we describe the specifics of the sky simulations we use; in Sect.~\ref{subsection:type_of_patches} we list the different types of patches and pixel subsets on which we perform the component separation independently; in Sects.~\ref{subsect:stat_syst_errors_on_betas}, \ref{subsubsection:Systematic residuals} and \ref{subsection:stat_sim_based} we perform the component separation with the different setups introduced and focus on the study of the foreground residuals in the cleaned CMB.
In Sect.~\ref{subsection:stat_semi_analytical} we validate the forecasting approach against the full simulation-based implementation, 
and finally, in Sect.~\ref{subsection:realistic_pixsets} we further discuss the trade-off statistical versus systematic residuals and a possible generalization of the definition of pixel subsets.

\subsection{Foreground models and sky simulations}
\label{subsection:sims}

To illustrate the component separation performance, we use the \texttt{PySM3} \texttt{d1s1} and \texttt{d10s5} foreground models introduced above, that are based on spatially varying spectral indices.
As a consistent choice given the models considered, we assume a modified blackbody and a power law as parametric scalings in the mixing matrix for the component separation.

For each instrument we simulate foregrounds (polarized dust and synchrotron emissions), CMB frequency maps, and a set of $100$ noise frequency maps.
We sum these contributions to get the full frequency map simulations.
Note that we do not take different realizations of the CMB, but always add in the frequency maps the CMB realization of the \texttt{c1} model of \texttt{PySM3} where the input value of the tensor-to-scalar ratio is $r=0$.
This is not a limiting factor for our study as the CMB itself does not affect the component separation performance: it does not leak outside of the reconstructed CMB as the CMB scaling is assumed to be known. 

We assume delta bandpasses and no beam convolution. 
Including the former is trivial and would not affect the conclusions of our paper. 
Including the latter instead is trivial if the same beam is assumed in all the frequency bands, and requires a more careful treatment if different frequency bands are convolved with different beams (see~\cite{rizzieri2025validating}), however this goes beyond the scope of this work where we are interested in the response of the method to different choices of patches and pixel subsets in handling the spatial variability of the scaling laws. 
For the same reason, we neglect other realistic effects that should be included in a full analysis, such as potentially more complex noise and the misestimation of the noise covariance, resolution and beam effects, realistic bandpasses and other known and unknown systematic effects relevant for real data, as well as foreground models with more complex scaling per pixel, either due to intrinsically more complex foreground properties or to effects arising from a pre-processing.
In particular, the latter point means that we simulate the frequency maps at the \texttt{nside}=64 resolution directly, which is a simplification with respect to simulating them at higher \texttt{nside} and then downgrading them, which would partially destroy the MBB and PL scalings of the foregrounds pixel-wise.

\subsection{Type of independent pixel subsets}
\label{subsection:type_of_patches}
\begin{figure}
    \centering
    \includegraphics[width=\columnwidth]{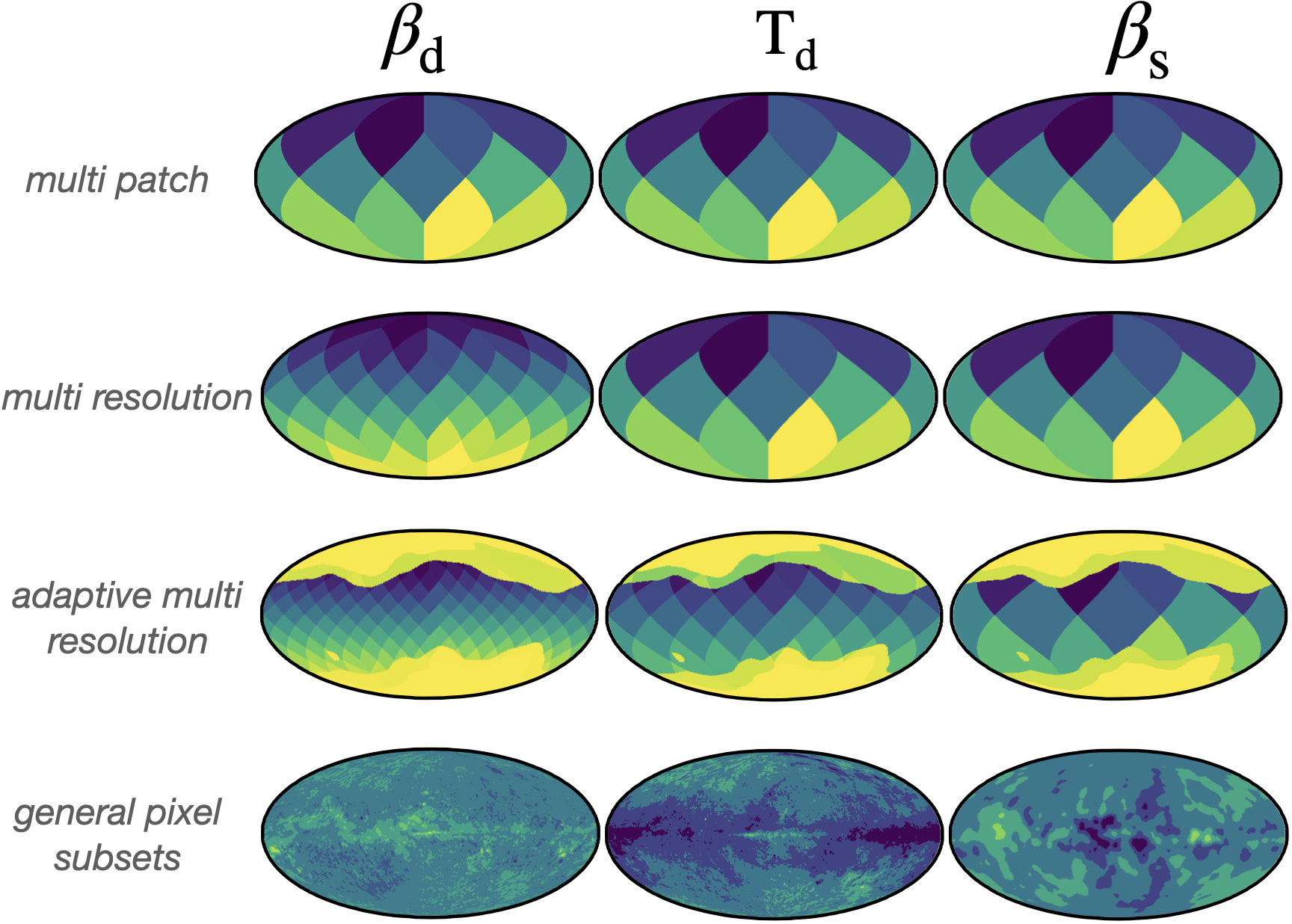}
    \caption{Illustration of the various definitions for the spatial variability of spectral parameters in the component separation, such as multi patch~\citep{errard2019characterizing} and adaptive multi resolution~\citep{litebird2023probing} or generic pixel subsets. Pixels with a same color belong to the same pixel subset.}
    \label{fig:multi_options}
\end{figure}

In the previous sections we have presented a general formalism and implementation allowing to associate to each pixel the pixel subset it belongs to for each spectral parameter type (in our case $\beta_\text{d}$, $\mathrm{T}_\text{d}$ or $\beta_\text{s}$) when performing the component separation.
We now give a few examples of how such pixel subsets can be chosen:
\begin{description}
    \item[\textbf{multi patch}:] the first, most straightforward, option could be to define them as super pixels, leveraging on the nested structure of the \textsc{HEALPix} scheme~\citep{gorski2005healpix}. We assume the spectral parameters to be constant over all the pixels that are inside a \textsc{HEALPix} pixel of given \nside\ parameter $q$~\citep{errard2019characterizing}.

    \item[\textbf{multi resolution}:] as different spectral parameters parameterize various sky components or different properties of those, we can extend on the previous approach by considering different $q$ for the different spectral parameters.

    \item[\textbf{adaptive multi resolution}:] for each spectral parameter the assumed $q$ varies in different patches of the sky (e.g.~with the sky latitude as the different regions shown in Fig.~\ref{fig:masks}). This approach was adopted to produce the forecast of LiteBIRD performance in~\citep{litebird2023probing}.
    
    \item[\textbf{general pixel subsets}:] for each spectral parameter each sky pixel can be assigned to any pixel subset, this should get as close as possible to the true underlying distribution of the spectral parameter template values. 
    There are two main aspects here that improve on the adaptive multi-resolution patches: first, going from the patches to generic shape pixel subsets, where two pixels far away in the sky can belong to the same subset; second, trying to represent the underlying morphology of the spectral parameter themselves, mimicking as much as possible the true underlying physical spectral parameter in the sky (or in the considered simulations).
    On simulations this can be easily done by binning the spectral parameter templates from which the model is built (called in the following ``ideal'' pixel subsets\footnote{Note that the term ``ideal'' here has been chosen to emphasize that these same pixel subsets, built with this procedure, could not be recovered in a real data scenario, since they rely on information derived from our perfect knowledge of the foreground models when dealing with simulations. We do not mean that they are ``ideal'' in the sense of ``optimal'' pixel subset configurations, although they are satisfactory for the cases studied in this work.}), while from the data themselves the challenge is to find a suitable quantity to bin. 
    Similar approaches have been investigated in the context of blind component separation techniques~\citep{khatri2019data, carones2023multiclustering}, those however are generically tracing the spatial variability of the foregrounds, and they are not trivially applicable here, as for the parametric component separation we need a tracer of the spatial variability of each spectral parameter\footnote{For instance binning a ratio of two input frequency maps as explored in those works for blind component separation would not be enough here mainly because it would not allow to disentangle between the spatial variability of \Bd\ and \Td.}.
    This feature makes the definition of the pixel subsets for the parametric component separation more challenging than for blind component separations, but on the other hand also more flexible, allowing to adapt to the specific morphology of the spatial distribution of each foreground SED (and their parameters).
    As a quantity to bin in a realistic scenario we propose in this work the spectral parameters recovered from a first component separation performed with multi-patch or multi-resolution configurations (``realistic'' pixel subset). This is however not a fully satisfactory tracer, and alternative options should be explored in the future, see Sect.~\ref{subsection:realistic_pixsets}.
    Note that in both the ``ideal'' and ``realistic'' pixel subset definitions we bin on the full sky with $n_\text{bins}$ equally spaced bins from the minimum to the maximum value present in the considered templates.
    Alternative, more sophisticated, binning procedures could be followed but we limit our study to this simple recipe as it is sufficient in the ideal case of binning the spectral parameter templates. 
    In addition, it appeared in the many tests we conducted that the definition of the binning itself has a minor influence on performance, which is instead more affected by the template over which the binning is performed.
    In the following we label the pixel subset cases with $n_\text{bins}$ in the full sky, which is always greater than or equal to the resulting number of pixel subsets in the fraction of sky observed.
\end{description}
Visual representations for all these cases are given in Fig.~\ref{fig:multi_options}.
The weakness of the multi-patch approach lies in its limited flexibility in choosing the patches, while the more flexible multi resolution and adaptive multi resolution present the problem of requiring the component separation user to make excessive choices in the fine tuning of the patch parameters, hardly fully justifiable\footnote{Even though some physical intuition could be behind those, or testing many configurations and looking at the recovered values on $r$ and its uncertainty could give some guidelines for the patch definition.} in a realistic scenario (unless built on simulated foreground and then used for the real data case).
However, in the following we still consider these configurations, as they are instructive for understanding the performance of the component separation and can serve as a basis for constructing realistic pixel subsets.
We note this is not a complete list of all the possible approaches to follow, and others may be pursued.
Examples with patches with more flexible shapes than \textsc{HEALPix} patches are suggested in~\cite{puglisi2022improved}, where the possibility to build such patches on external data is also explored.

\subsection{Statistical and systematic errors on the spectral parameters}
\label{subsect:stat_syst_errors_on_betas}
\begin{figure}
    \centering
    \includegraphics[width=\linewidth]{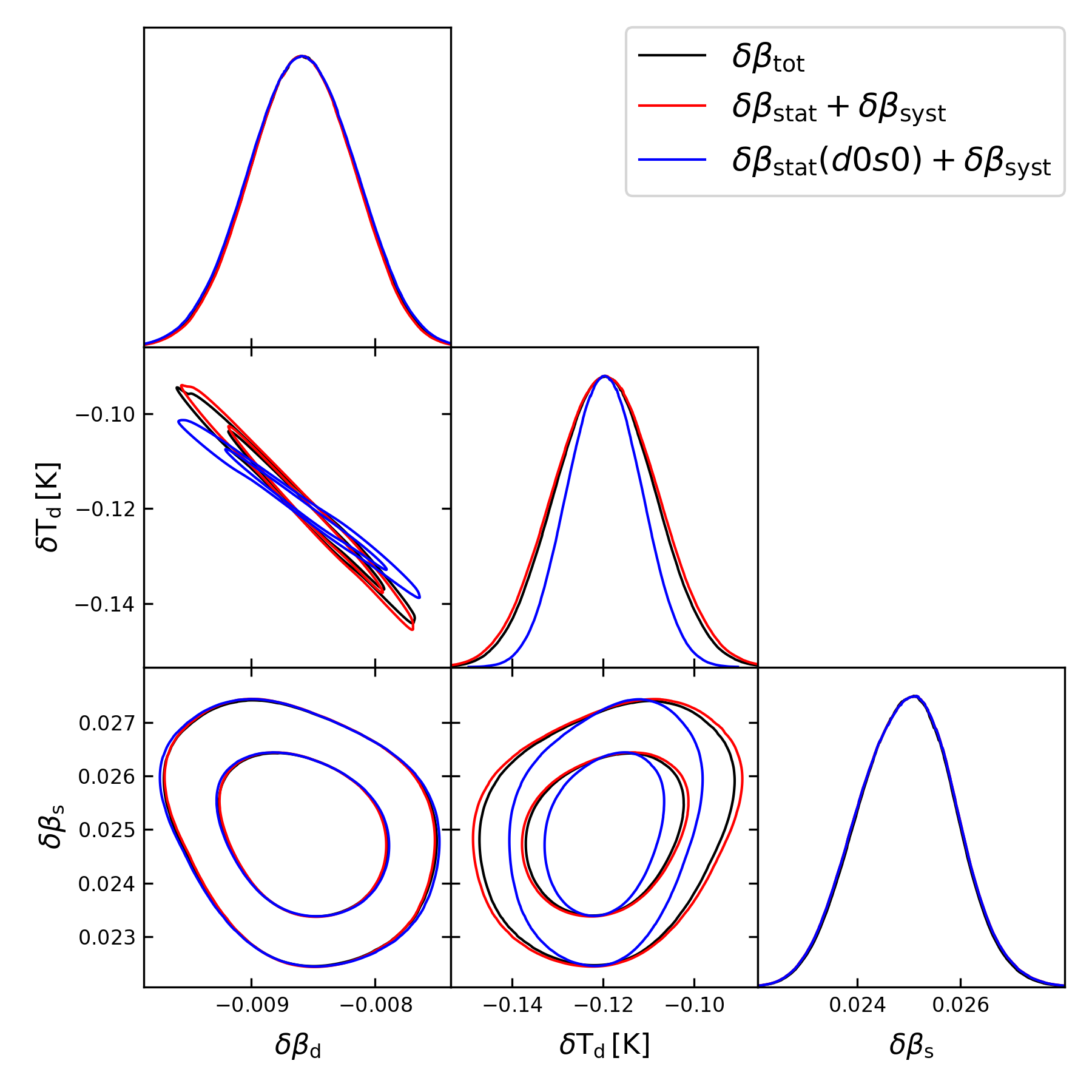}
    \caption{
    For the PICO-like setup, with foreground model \texttt{d1s1} and \texttt{ns}=2 patches, we show the error on the recovered spectral parameters as computed from the total simulations ($\delta \bm \beta_{\mathrm{tot}}$) versus computed by summing the estimation of the spectral parameter errors including the systematic error only ($ \delta \bm \beta_{\mathrm{syst}}$) and the statistical error only ($\delta \bm \beta_{\mathrm{syst}}$). 
    We use two different methods to compute the statistical residuals (see the main text) to show that those depend on the foreground model considered. This dependence is particularly visible on \Td.}
    \label{fig:sum_betas}
\end{figure}

We verify on simulations that the error on the recovered spectral parameters $\delta \bm{\beta}_{\mathrm{tot}}$ can indeed be decomposed into a statistical and a systematic contribution (respectively, $\delta \bm{\beta}_{\rm stat}$ and $\delta \bm{\beta}_{\rm syst}$), and we suggest approximate ways to compute these two contributions independently on simulations.
We want hence to verify that in a given pixel, $\delta \bm{\beta}_{\mathrm{tot}}=\delta \bm{\beta}_{\rm stat}+\delta \bm{\beta}_{\rm syst}$ holds.
In the case of the \texttt{d1s1} sky model, we show in Fig.~\ref{fig:sum_betas} both $\delta \bm \beta_{\mathrm{tot}}$ and $\delta \bm \beta_{\mathrm{stat}} + \delta \bm \beta_{\mathrm{syst}}$ as found for a given pixel from the component separation over 100 noise realizations.
We compute $\delta \bm \beta_{\mathrm{tot}}$ as the recovered spectral parameter values after spectral likelihood maximization, to which we subtract the true values for that specific pixel.
Instead, we compute $\delta \bm \beta_{\mathrm{syst}}$ as the recovered spectral parameters from a component separation run on noiseless frequency maps $\mathbf{\hat{d}}$ to which we subtract the true spectral parameter values.
In practice this means maximizing Eq.~\eqref{eq:avg_spec_like} where $\mathbf{\hat{d}}$ instead of $\mathbf{d}$ ensures that no noise intervenes in the identification of the maximum of the likelihood and that any biases in the scaling law assumptions is the only effect affecting the estimated spectral parameters.
It is not possible instead to access exactly the $\delta \bm \beta_{\mathrm{stat}}$ term, as the spectral likelihood (Eq.~\eqref{eq:spec_like_pixel_sum}) depends on the foreground sky present in the frequency map simulations $\mathbf{d}$ which inevitably come with a bias due to the foreground complexities.
We can however estimate it approximately as the component separation result on simulations with a customized model which has the same scaling laws as the ones assumed in the mixing matrix of the component separation (so that no bias due to the scaling law assumptions is introduced) to which we subtract the true spectral parameter values of this customized model for the considered pixel.
To build such a customized model we average the true spectral parameter values of the initial foreground model present in the simulations in each pixel subset.
To further emphasize the dependence of $\delta \bm \beta_{\mathrm{stat}}$ on the sky model and the fact that our customized model is a good approximation, we also estimate it using simulations with the \texttt{d0s0} model.

In the cases with higher number of pixel subsets for the LB-like scenario, because of the low S/N to reconstruct the spectral parameters in each pixel subset, it is not possible to set strong constraints on them. The distribution of the recovered spectral parameter values per pixel from the different noise realizations would hence be very wide.
To avoid going to nonphysical values we hence set bounds in the minimization of $\beta_{\rm d} \in \left[0, 5 \right]$, $\rm{T}_{\rm d} \in \left[10, 40 \right]$K and $\beta_{\rm s} \in \left[-6, 0 \right]$.
Alternatively, more well-behaved options such as Gaussian priors could be considered. 
However, we do not explore these alternatives in this work, as such bounds only become relevant for cases (such as the \texttt{ns16} or the \texttt{ns32} shown in the following) that we include in our plots in the following to illustrate the behaviour of the residuals, but would not be suitable for a cosmological analysis, where configurations with a lower number of patches should be preferred.

\subsection{Systematic residuals}
\label{subsubsection:Systematic residuals}
\begin{figure*}
    \centering

    \begin{minipage}[b]{0.48\textwidth}
        \centering
        \includegraphics[width=\linewidth]{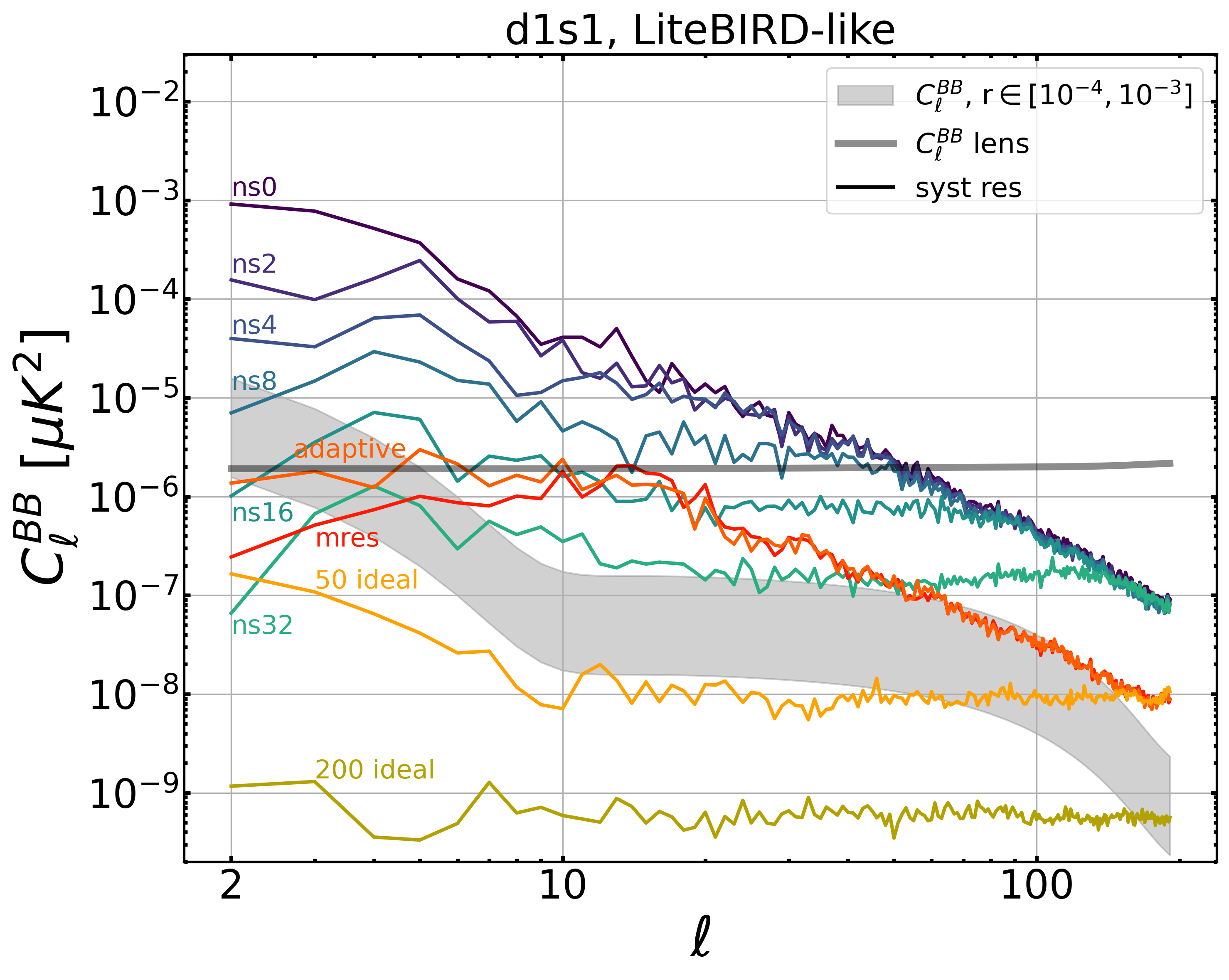}
    \end{minipage}
    \hspace{0.02\textwidth}
    \begin{minipage}[b]{0.48\textwidth}
        \centering
        \includegraphics[width=\linewidth]{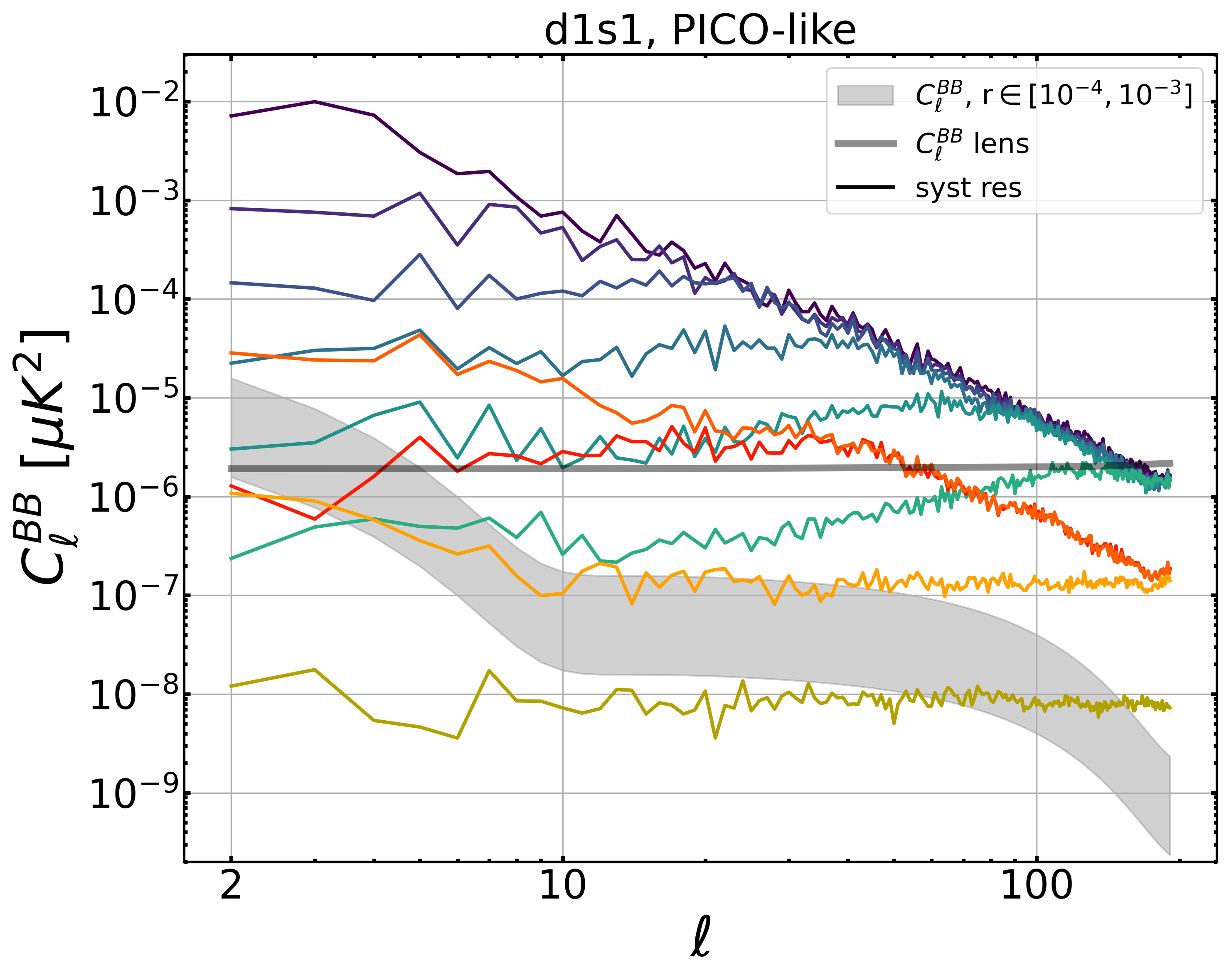}
    \end{minipage}

    \begin{minipage}[b]{0.48\textwidth}
        \centering
        \includegraphics[width=\linewidth]{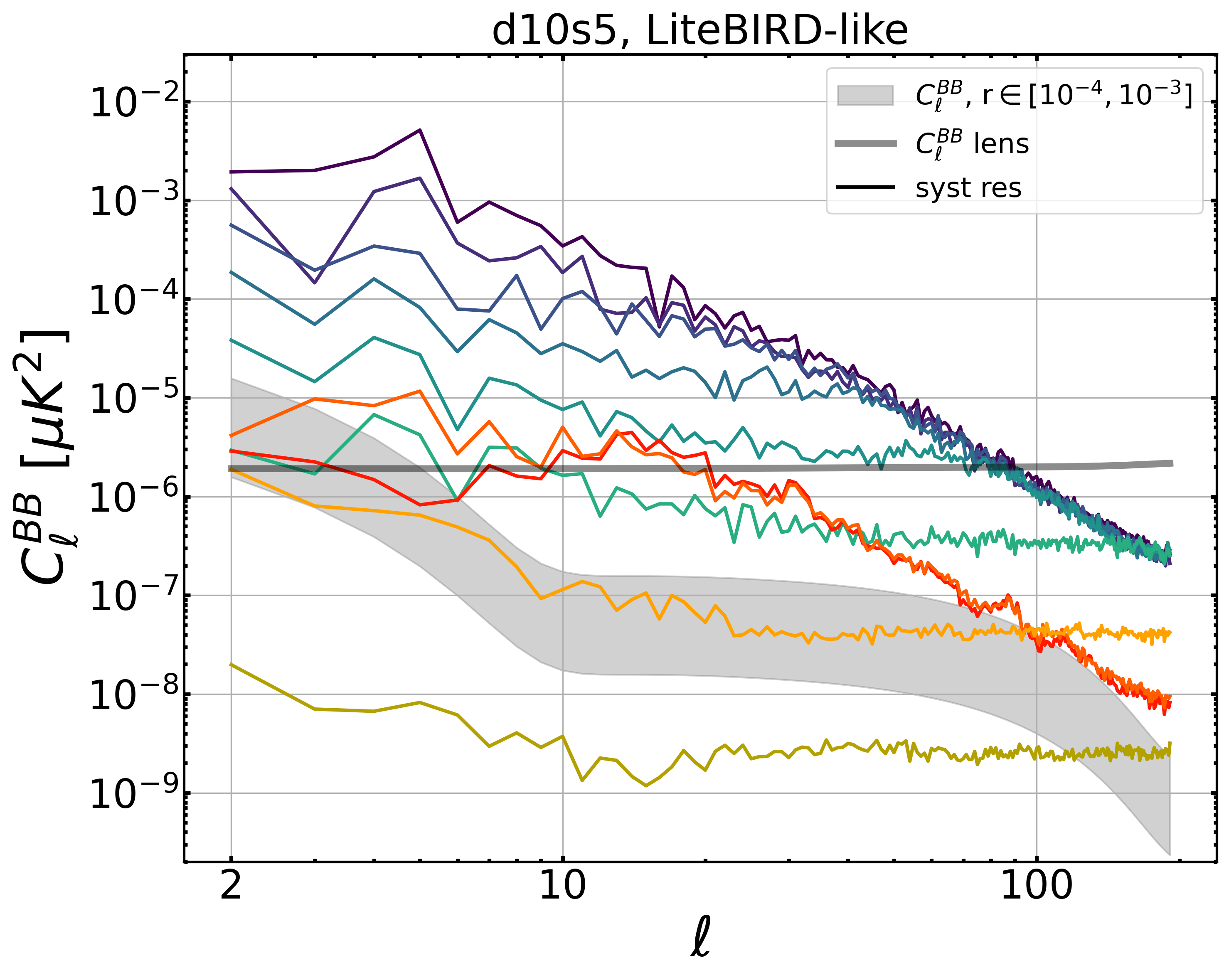}
    \end{minipage}
    \hspace{0.02\textwidth}
    \begin{minipage}[b]{0.48\textwidth}
        \centering
        \includegraphics[width=\linewidth]{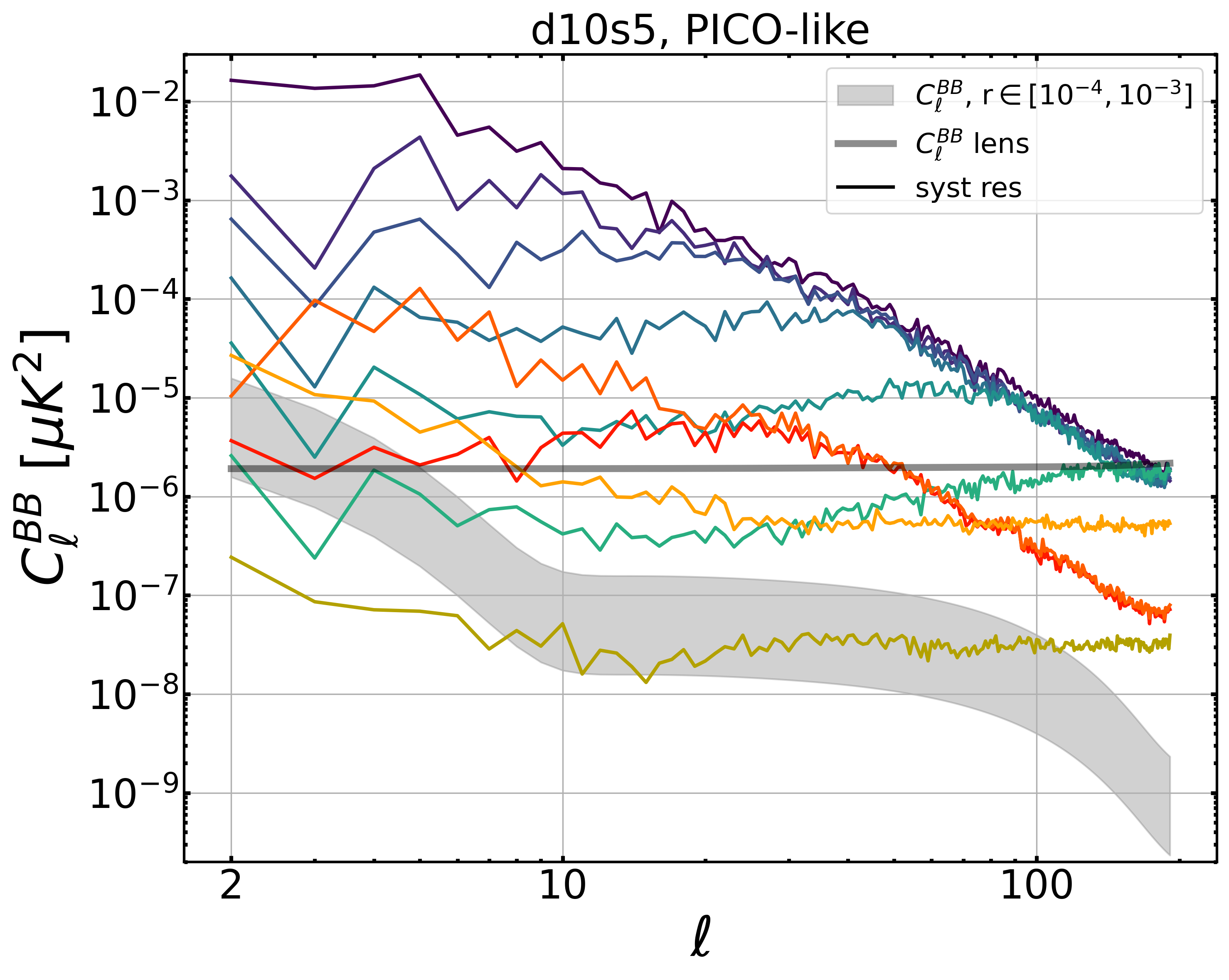}
    \end{minipage}

    \caption{Example of power spectra of the systematic residuals for \textit{LiteBIRD}-like and \textit{PICO}-like setups, \texttt{d1s1} and \texttt{d10s5} sky models, for fsky=60\%, and various pixel subset configurations.
    The configurations with `nsX' are \textsc{HEALPix} multi-patch configurations at the specified \texttt{nside}, `mres' corresponds to a \textsc{HEALPix} multi-resolution configuration with \texttt{nside}=[64,8,4] respectively for 
    [$\beta_\text{d}$, $\mathrm{T}_\text{d}$, $\beta_\text{s}$] and 
    `adaptive' the configuration 
    of~\protect\cite{litebird2023probing}, 
    that has the same number of patches of `mres' at low Galactic latitude and slightly less at high Galactic latitudes.
    The residuals are typically lower the higher the number of pixel subsets considered for \textsc{HEALPix} patches. With the `ideal' pixel subsets instead we find configurations with low number of patches giving negligible systematic residuals.
    Taking as many pixel subsets as pixels (corresponding to multi patch with ns64) would result in numerically zero systematic residuals, as the underlying scalings per pixel in the simulations are a modified blackbody and a power law, as assumed in the mixing matrix.
    }
    \label{fig:minipage2x2_systematics}
\end{figure*}

In the different configuration of pixel subsets proposed above we look at the systematic foreground residuals, for the LB-like and PICO-like scenarios, for different pixel subsets configurations, see Fig.~\ref{fig:minipage2x2_systematics}.
Each panel shows, for an instrument configuration and a foreground model, the angular power spectra of the systematic foreground residual maps, in $\mu {\rm K}^2$, as a function of the multipole $\ell$. 
The spectra are compared to the theoretical primordial $B$ modes for $r$ ranging from $10^{-4}$ to $10^{-3}$ as well as the lensing $B$ modes for $A_{\rm lens}=1$.

\begin{figure}
    \centering
    \includegraphics[width=\columnwidth]{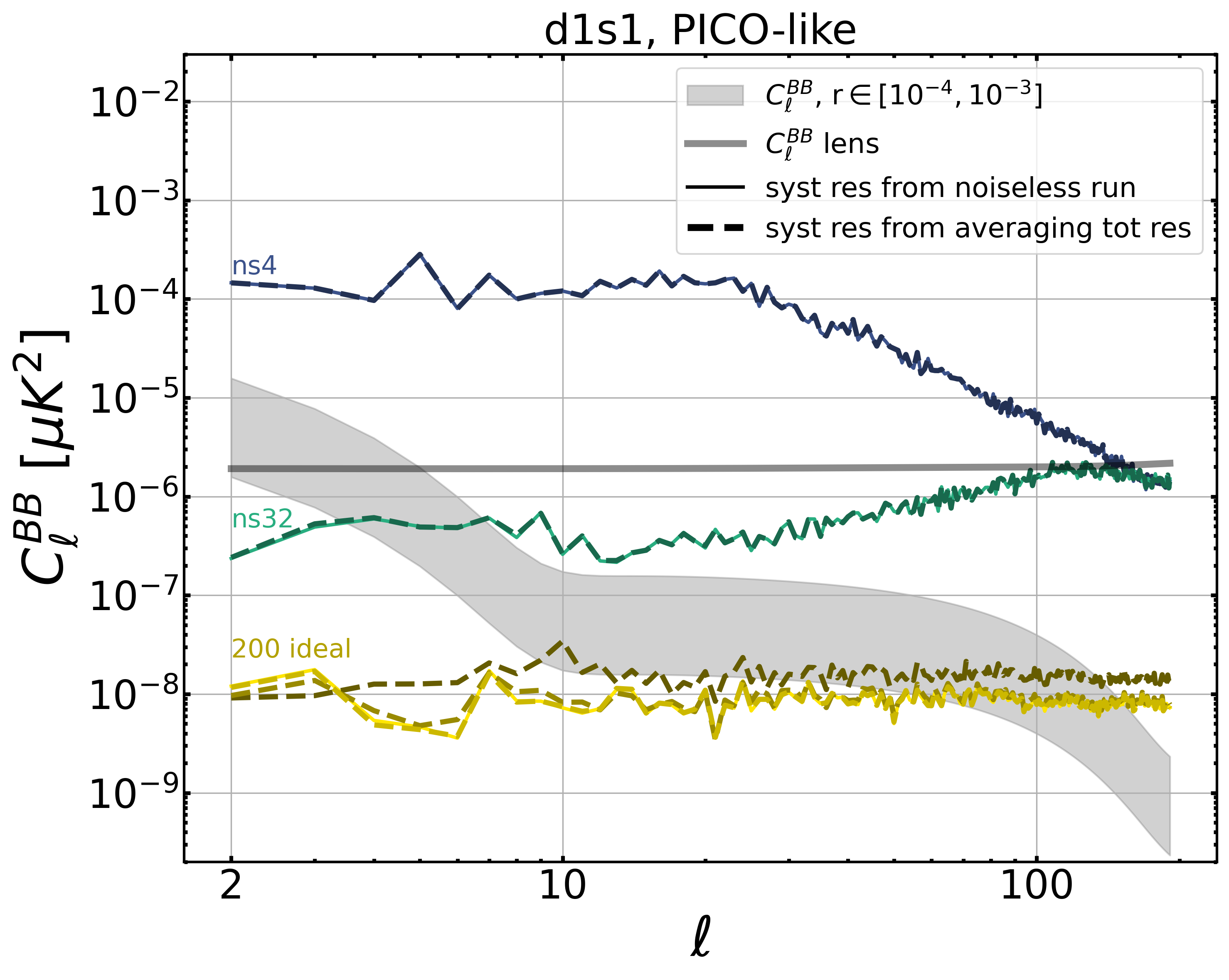}
    \caption{Comparing the systematic residuals computed from a noiseless component separation run or by averaging the residuals from the full component separation from 100 realizations. The latter, for the ideal case, is shown for different numbers of realizations considered (10, 100 and 1000 for the darker to lighter dashed lines). Because of the low level of the systematic residuals, the error from averaging on a low number of configurations is visible by eye, however this is reduced increasing the number of simulations.}
    \label{fig:syst_res_from_noiseless_and_from_average}
\end{figure}

Assuming $\delta \mathbf{s}_{\mathrm{syst}}$ depends on $\delta \bm \beta_{\mathrm{syst}}$ only, on simulations, $\delta \mathbf{s}_{\rm syst}$ can be estimated from a run of the component separation on noiseless foreground maps: the obtained residuals are the irreducible leakage of foregrounds into the recovered CMB map, due to the mis-modeling of the ``true'' foreground emissions. 
We then take the power spectrum of these maps, $C_{\ell}^{\mathrm{syst}}$\footnote{Note that here and in the following we compute the power spectra with the \texttt{anafast} routine of \texttt{healpix}. We have verified for a subset of spectra presented in this work that using \texttt{NaMaster} the spectra do not change. This is the case because we always compute them for a large sky fraction and for the CMB residuals only, hence avoiding $E$-to-$B$ leakage issues.}. 
This is a computationally more convenient approach than taking the average of the full component separation residuals.
The latter tends to the noiseless component separation result as the number of simulations we average over increases. We show this in Fig.~\ref{fig:syst_res_from_noiseless_and_from_average} for a limited number of cases.
We find similar results whenever we consider cases with sufficient S/N so that the bounds we have set are not hit. 
On the contrary, we expect differences when the bounds are reached. 
In the following, we adopt the estimate of systematic residuals from a noiseless run which turns out to be the same procedure to calculate them as in the semi-analytical forecasting implementation.

For the \textsc{HEALPix} patches the trend is that for more patches the systematic residuals decrease, as the spatial variability of the foregrounds is better and better accounted for.
However, only a limited number (tens or few hundreds) of pixel subsets could give negligible systematic residuals if those well represent the underlying spectral parameter templates.
With the multi-resolution configuration (with \texttt{nside}=[64,8,4] respectively for 
    [$\beta_\text{d}$, $\mathrm{T}_\text{d}$, $\beta_\text{s}$]) we see that it is helpful to consider different patch resolutions for the different spectral parameters, as the spatial variability morphology of the different spectral parameters also differs.

From comparing the resulting systematic residuals for the two foreground scenarios studied as well as for the two instrumental configurations considered, we see that the level of systematic residuals recovered depends on the actual \texttt{PySM3} templates but also on the instrumental frequency range (in addition to the chosen sky area for the analysis, as the foregrounds are typically more complex closer to the Galactic plane).
Indeed, as the systematic foreground residuals are estimated from noiseless frequency maps, they do not depend on the noise in the observed data\footnote{However, the systematic residuals depend mildly on the noise covariance assumed for the component separation, not necessarily equal to that of the noise in the data, which may become important when the noise covariance is mis-estimated.
}. 
A perfect observatory, with unlimited sensitivity, could suffer from this bias if there is a mismatch between the underlying foreground SEDs and the models in the mixing matrix.
In particular, the poorer performance of the PICO-like case when having the same foreground model and number of patches is due to its broader frequency coverage. 
This wider coverage makes it more sensitive to spatial variations in the \Td\ template, whereas the LB-like case, with its lower maximal frequency band, is less affected by these variations.

\subsection{Statistical residuals (simulation-based)}
\label{subsection:stat_sim_based}
The complementary quantity when evaluating the performance of a parametric component separation are the statistical foreground residuals.
In the notation of the previous sections this term is $\delta \mathbf{s}_{\mathrm{stat}}$.
Those, together with the noise present in the frequency maps and remaining in the reconstructed CMB map give the statistical contribution to the residuals.
We compute them by subtracting the systematic map we have computed in the previous section from the recovered total residuals map.
Hence, by definition, these residuals are non-zero for a given foreground+noise realization, but their maps will average to zero when averaging over noise simulations. Their RMS, or angular power spectrum, will not be zero.
Once we have the difference maps we take their spectra, and average those.
We focus here on statistical residuals computed through simulations as just described, while we discuss in Sect.~\ref{subsection:stat_semi_analytical} the statistical residuals computed with the semi-analytical implementation and the comparison between these two cases.
\begin{figure*}
    \centering
    \begin{minipage}[b]{0.48\textwidth}
        \centering
        \includegraphics[width=\linewidth]{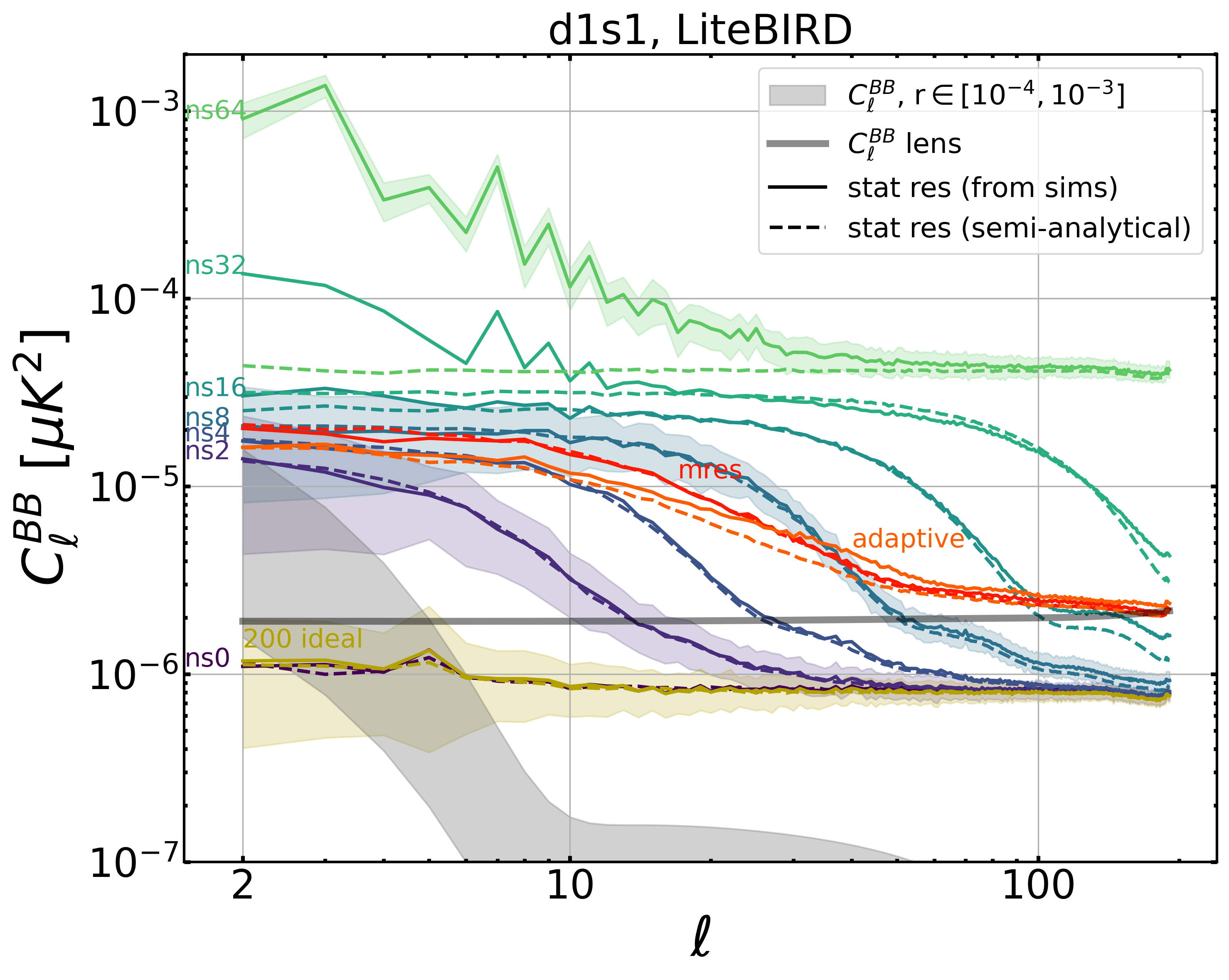}
    \end{minipage}
    \hspace{0.02\textwidth}
    \begin{minipage}[b]{0.48\textwidth}
        \centering
        \includegraphics[width=\linewidth]{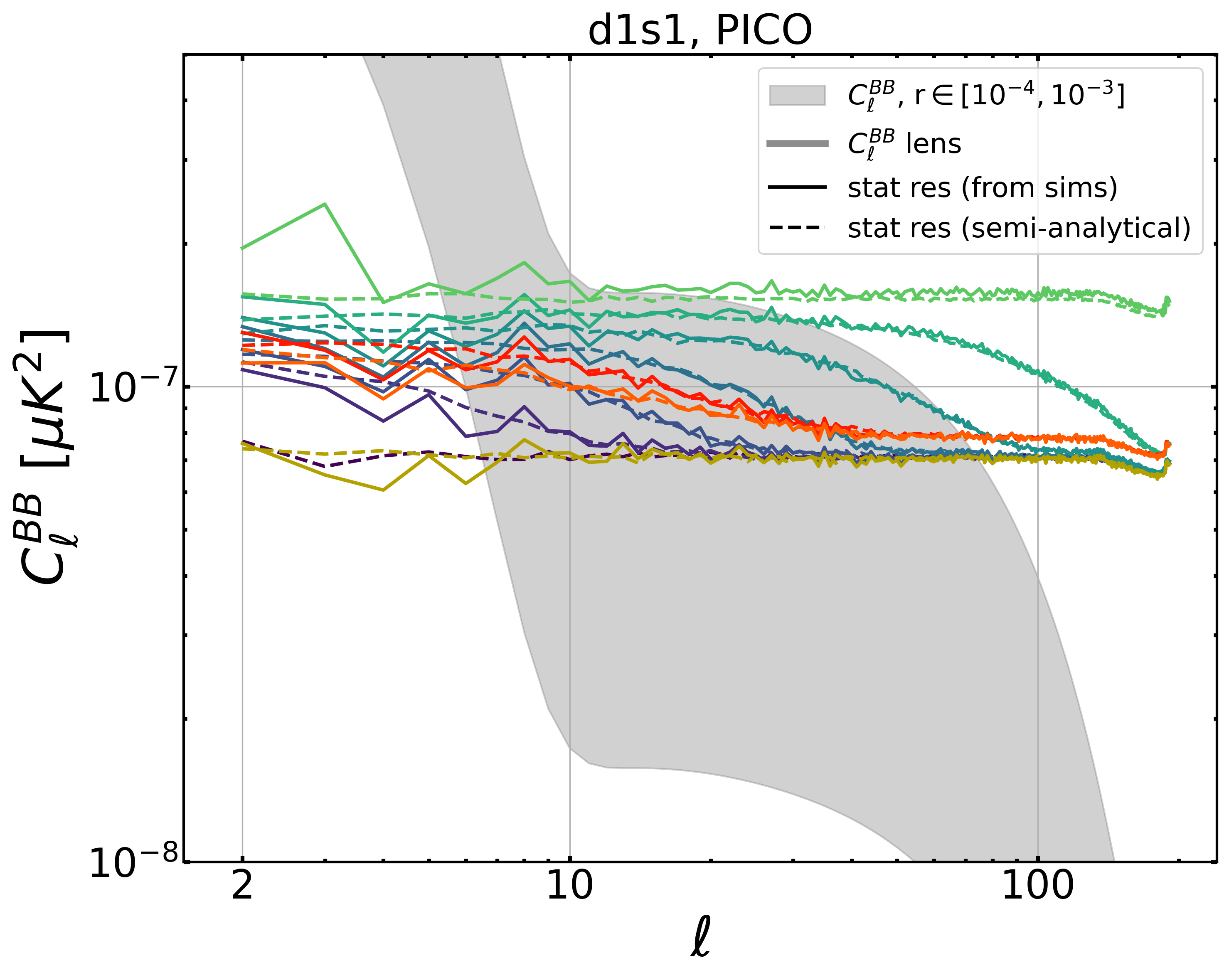}
    \end{minipage}
    \caption{Mean of the power spectra of the foreground statistical residuals for the two missions \textit{LiteBIRD}-like and \textit{PICO}-like, for the same pixel subsets configurations of Fig~\ref{fig:minipage2x2_systematics}.
    Computed both with the simulation-based implementation (100 simulations) and with the semi-analytical formalism of \texttt{xforecast} (500 simulations).
    For some of the LB-like simulation-based cases we also show the one standard deviation range (for clarity of the plot we avoid plotting this for the PICO-case).
    Not to make the plot too busy we do not show the residuals for the cases of Fig.~\ref{fig:minipage2x2_systematics} with 50 ideal pixel subsets, which would be just below the 200 ideal line, with no appreciable difference by eye.
    Finally, notice that for visual purposes we set the y-axis to different scales.
    }
    \label{fig:stat_full_vs_xforecast}
\end{figure*}

In Fig.~\ref{fig:stat_full_vs_xforecast}, we show the mean of the power spectra of the statistical residuals ($C_{\ell}^{\mathrm{stat \; fg}}$) computed from the frequency maps with the 100 different noise realizations, for the \texttt{d1s1} cases.
In the LB-like case we also show the one standard deviation range from these realizations. 
We do not include them for the PICO-like setup to preserve the clarity of the figure.

The trend that we find here for pixel subsets built with a same procedure (for example the \textsc{HEALPix} patches at different \texttt{nside}, or the ideal pixel subsets with different number of total subsets)  is opposite to the one we had for the systematic residuals: the configurations with higher statistical residuals in Fig.~\ref{fig:stat_full_vs_xforecast} are the ones with lower systematic residuals in Fig.~\ref{fig:minipage2x2_systematics}.
More generally, the statistical residuals go up when increasing the number of patches.
It is interesting to notice however that this is strictly true when considering the patches per spectral parameter type, and not the overall number of patches.
The total number of patches is indeed lower for the case with \texttt{nside}=16, than the multi-resolution case (with \texttt{nsides}=[64,8,4] respectively for [$\beta_\text{d}$, $\mathrm{T}_\text{d}$, $\beta_\text{s}$]), however the latter gives lower statistical residuals than the former.
This suggests that not all spectral parameter types contribute equally to the statistical residuals. Moreover, having fewer pixel subsets for one parameter type may also help reduce the statistical residuals of other parameter types, because of the correlations between the different spectral parameters.
We can use the semi-analytical forecasting implementation to further investigate these hypotheses (see Sect.~\ref{subsection:stat_semi_analytical}).

For a given instrument, the statistical residuals we find for the two different foreground models studied (\texttt{d1s1} and \texttt{d10s5}) are very similar, contrary to the systematic residuals that were instead presenting very clear dependence on the foreground model specifics.
For this reason we do not show the \texttt{d10s5} statistical residuals, however
a mild dependence on the foreground model is present, we show it in the semi-analytical approach and discuss the implications of neglecting it in Sect.~\ref{subsection:stat_semi_analytical}.

Finally, comparing the two setups studied we find lower statistical residuals for the PICO-like setup (contrary to the systematic residuals that we showed in the previous section to be higher in the PICO-like setup). 
This is due to the much higher sensitivity of this setup, but also because of its broader frequency range which partially lifts the degeneracy between \Bd and \Td, reducing the uncertainty on these parameters.
We note that because of the different characteristics of the instruments and different performances of statistical and systematic residuals a particular pixel subset configuration optimized for one instrument is not necessarily optimal for another instrument.
This can be the case for example of the multi-resolution configuration optimized for the LB-like setup and not being the most performative for the PICO-like setup.

\subsection{Comparison simulation-based vs semi-analytical forecasting}
\label{subsection:stat_semi_analytical}
We now focus on getting the statistical residual estimates with the semi-analytical approach.
The semi-analytical implementation is far less computationally expensive than the simulation-based estimate when generating multiple realizations. 
This is because it avoids re-optimizing the spectral likelihood for each case. 
Instead, it only requires generating the desired number of random spectral parameter errors with the properties of Eq.~\eqref{eq:stat_prop_delta_beta}, performing a linear matrix product, and computing the corresponding power spectrum.
In this case, we hence produce 500 realizations of statistical residuals.

Note that in practice the Cholesky decomposition of the Fisher-like matrix can be numerically performed only if the matrix is well-conditioned.
We find this to be the case in all instances except for the ideal pixel subset cases in the full sky. 
However, masking the Galactic plane, as we would anyway do to avoid the most foreground-contaminated regions, is sufficient to ensure that the Hessian remains well-behaved even in these cases.

Along with the simulation-based statistical residuals in Fig.~\ref{fig:stat_full_vs_xforecast} we show the statistical residuals computed with the semi-analytical forecasting formalism.
We find overall good agreement between the two cases, with both the mean of the statistical residuals and their standard deviation ranges (only plotted for a few cases) overlapping for most pixel subsets and across most angular scales.

\begin{figure}
    \centering
    \includegraphics[width=\columnwidth]{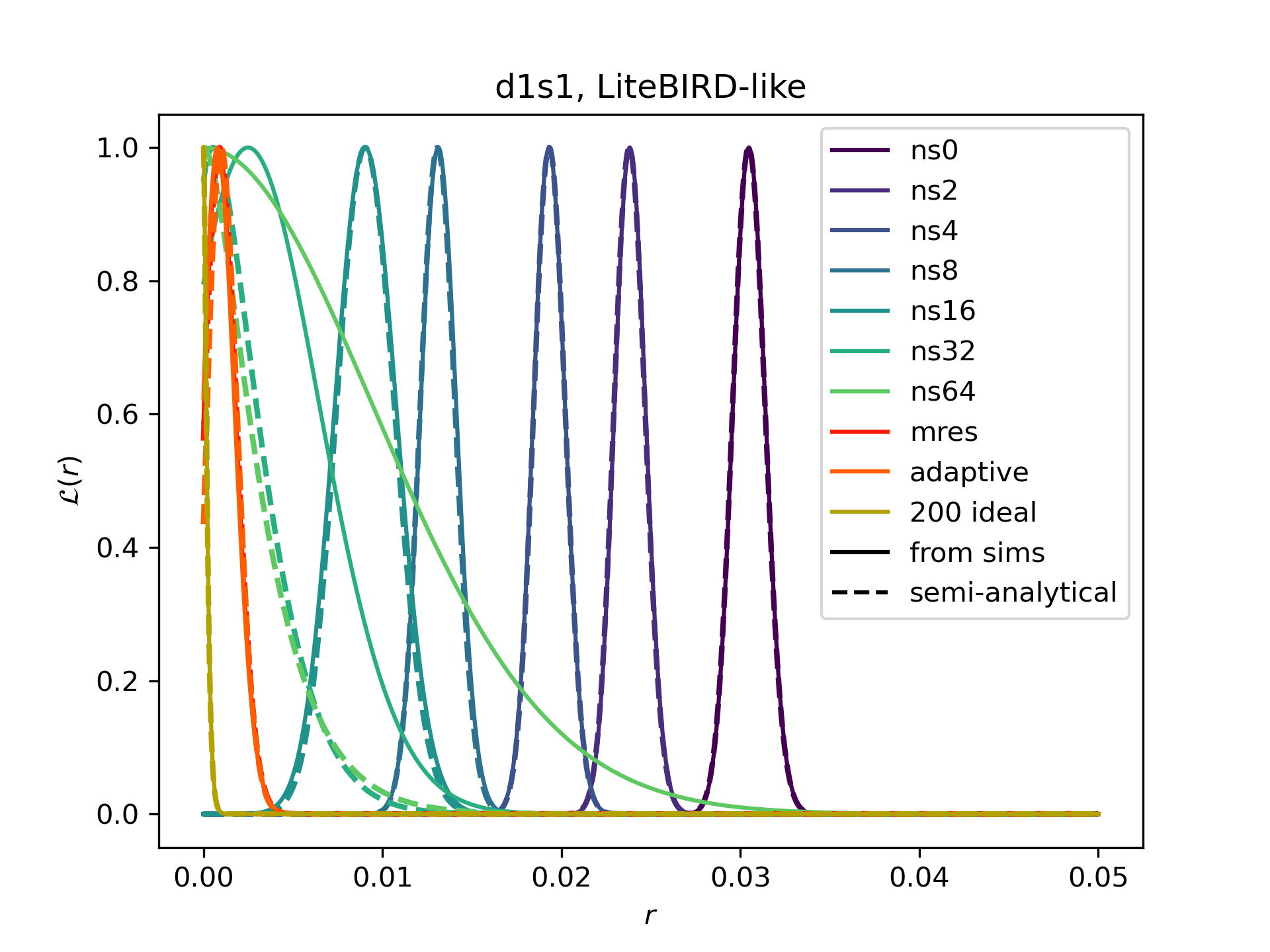}
    \caption{Posterior on $r$ computed with the statistical residual estimates from the simulation based and the semi-analytical implementations, for the LB-like scenario with \texttt{d1s1} foreground models.}
    \label{fig:posterior}
\end{figure}

To more quantitatively compare the residual estimates from the two implementations, we extend the forecasting to $r$ and examine the posterior obtained in each case.
To have a quick estimate of $r$ and $\sigma_r$ we use the following likelihood:
\begin{equation}
    \centering
        -2\log \left( \mathcal{L}(r) \right) = \sum_\ell{ \left(2\ell+1\right)f_{\rm sky}\left[\frac{C_\ell^{\rm obs}}{C_\ell^{\rm model}}+\log(C_\ell^{\rm model})\right]}
        \label{eq:cosmo_lik}
\end{equation}
where
\begin{align}
    &C_\ell^{\rm obs} \equiv C_\ell^{\rm BB,\,theory} + C_\ell^{\rm tot},\\
    &C_\ell^{\rm model} \equiv C_\ell^{\rm BB,\,theory} + C_\ell^{\rm stat}.
\end{align}
where, $C_\ell^{\rm tot}$ is the total residuals and in the semi-analytical forecasting formalism we take:
\begin{equation}
    C_\ell^{\rm tot} = C_\ell^{\rm syst} + C_\ell^{\rm stat}.
\end{equation}
From the peak of this likelihood we get an estimate for $r$.
We will refer to this as $r_{\rm bias}$ as the input value of $r$ in the simulations is 0.
We compute the uncertainty on the measured $r$, $\sigma_r$ as the 68\% confidence level from the peak of the likelihood.
We show the posterior we recover for the LB-like case in Fig.~\ref{fig:posterior} and give in Tab.~\ref{tab:r_values} the values recovered for $r_{\rm bias}$ and $\sigma_r$ for some of the pixel subset configurations (simulation-based and semi-analytical cases).
\begin{table}
\begin{tabular}{|c|c|c|c|}
\hline
 & sim-based ($\times 10^{-3}$) & semi-analytical ($\times 10^{-3}$) & $n_{\sigma}$ \\
\hline
\multicolumn{4}{|c|}{--- LB-like, \texttt{d1s1} ---} \\
adaptive   & $0.8 \pm 1$ & $1 \pm 1$ & 0.1 
\\
200 ideal  & $0.0004 \pm 0.2 $ & $0.0005 \pm 0.2 $ & 0.0005 \\
\hline
\multicolumn{4}{|c|}{--- LB-like, \texttt{d10s5} ---} \\
adaptive   & $3 \pm 1$ & $3 \pm 1$ & 0.06 \\
200 ideal  & $0.003 \pm 0.2$ & $0.003 \pm 0.2$  & 0.00003 \\
\hline
\multicolumn{4}{|c|}{--- PICO-like, \texttt{d1s1} ---} \\
adaptive   & $23 \pm 0.7$ & $23 \pm 0.7$ & 0.003 \\
200 ideal  & $0.006 \pm 0.1$ & $0.007 \pm 0.1$ & 0.001 
\\
\hline
\multicolumn{4}{|c|}{--- PICO-like, \texttt{d10s5} ---} \\
adaptive   & $20 \pm 0.7$ & $20 \pm 0.7$ & 0.003\\
200 ideal  & $0.04 \pm 0.2$ & $0.04 \pm 0.2$ & 0.003\\
\hline
\end{tabular}
\caption{We report the recovered values of $r_{\mathrm{bias}}\pm \sigma_r$ for two of the configurations studied, both when using the simulation-based residuals and the ones from the semi-analytical implementation. As the CMB simulations used have input $r=0$, the $r$ value we find as the maximum of the likelihood of Eq.~\ref{eq:cosmo_lik} corresponds to the bias due to the component separation residuals. $\sigma_r$ is instead the $68\%$ c.l. from the maximum from the posteriors shown in Fig.~\ref{fig:posterior}.
As a measure of the compatibility of the two results we give the number of $\sigma$ by which those differ, computed as the ratio of the absolute value of the difference of $r_{\mathrm{bias}}$ in the two cases and sum in quadrature of their $\sigma_r$.}
    \label{tab:r_values}
\end{table}
This comparison thus shows that the neglected cross terms in summing at the power spectrum level in the semi-analytical approach, as well as the 2nd order term of Eq.~\eqref{eq:residual_map}, are negligible.
Also, the difference visible for the intermediate scales in the statistical residuals of \texttt{d1s1} LB-like does not significantly impact the final values of $r$, as those are still compatible to well below 1 sigma.

\begin{figure*}
    \centering
    \includegraphics[width=\linewidth]{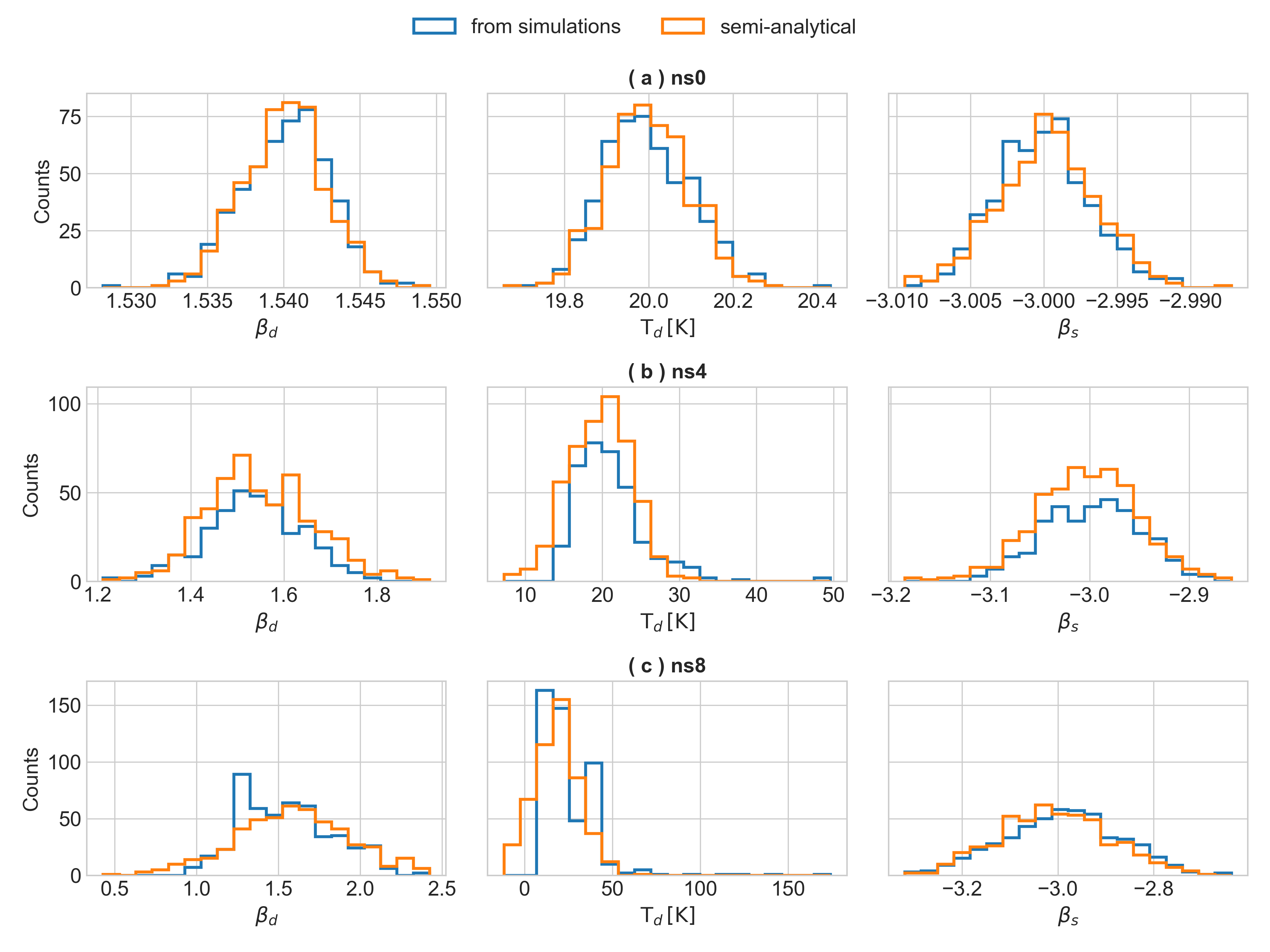}
    \caption{For the \textit{LiteBIRD}-like case we plot the distributions of the recovered spectral parameter values from 500 sky simulations, with the \texttt{d0s0} foreground model and different white noise realizations.
    This is done both for the spectral parameters as found from the simulation-based approach and with the semi-analytical implementation. In the upper plot this is shown for the case where a single spectral parameter value is recovered from the full available sky area. In this case we are in a regime of high S/N and the two distributions are overlapping.
    Going to the case where the sky is divided in more and more subsets of pixels the S/N in each of those decreases.
    In those cases the distributions from the semi-analytical cases are still Gaussian, while wider than the \texttt{ns0} case.
    The ones from the simulation cases for \Td\ present a tail at high values, which in turn impacts the distribution of \Bd\ as those are correlated parameters.
    Note that the x-axis for the \texttt{ns8} case for \Td\ has been limited for visualization purposes, but the tail extends to $~10^5$~K.}
    \label{fig:beta_distribution}
\end{figure*}

A notable exception is however at large scales as the number of patches increases, where the simulation-based statistical residuals have more power. 
This starts to be noticeable at the \texttt{ns16} case and becomes more visible for the \texttt{ns32} and \texttt{ns64} cases, where a difference in the posteriors computed with the two implementations is also well visible.
The discrepancy hence becomes increasingly more important as the number of parameters increases and it is more relevant in the LB-like case where the S/N is lower.
Such a discrepancy between the simulation-based and semi-analytical method arises from the assumption of Gaussian distributed spectral parameters.
We show in Fig.~\ref{fig:beta_distribution} the distribution of the reconstructed spectral parameters, for a given pixel, from 500 noise realizations for the \texttt{ns0} and the \texttt{ns8} case, where this effect starts to be visible, in the LB-like scenario.
We over plot the distributions of the spectral parameters as found both from the simulation-based approach and the semi-analytical implementation if we do not set bounds for the recovered values of the spectral parameters.
We show this on the \texttt{d0s0} model, so that we can compare the spectral parameter values recovered with the \texttt{ns0} and \texttt{ns16} patch cases.
In the \texttt{ns4} and \texttt{ns8} cases of the figure the skewness is already visible for \Td\ and large values are present even if it only includes a limited number of cases among the 500 realizations, which means that when looking at a sky map only a few pixels will correspond to values in this tail.
The discrepancy between the simulation based and the semi-analytical implementations is thus not visible in these cases at the power spectrum level, but becomes more important for the \texttt{ns16}, \texttt{ns32} and \texttt{ns64} cases.
Moreover the residuals we are showing in Fig.~\ref{fig:stat_full_vs_xforecast} for the simulation-based case do not directly come from the spectral parameter distributions of Fig.~\ref{fig:beta_distribution} as bounds are applied during the maximization/minimization procedure.
In the semi-analytical approach, to mimic the same bounds of the simulation-based approach, we could constrain the $\delta \bm \beta$ random realizations that we create with the properties of Eq.~\eqref{eq:stat_prop_delta_beta}.
This can be easily done, however it does not correspond exactly to having the bounds in the minimization because the spectral parameters are correlated.
In addition, robustly including the bounds in the semi-analytical approach would be irrelevant in practice, since the bounds are reached only in regimes with low S/N per pixel subset -- for example, when fitting for different spectral parameter values in each pixel or nearly in each pixel, which is exactly what we aim to avoid by appropriately choosing the pixel subsets.
In conclusion, this discrepancy identifies a limit of validity of the semi-analytical method to reproduce the statistical residuals.
Still we find this difference not to be a strong limitation as the disagreement only concerns cases that have too high levels of statistical residuals and would not be suitable pixel subset configurations for a cosmological analysis.

\begin{figure*}
    \centering
    \begin{minipage}[b]{0.48\textwidth}
        \centering
        \includegraphics[width=\linewidth]{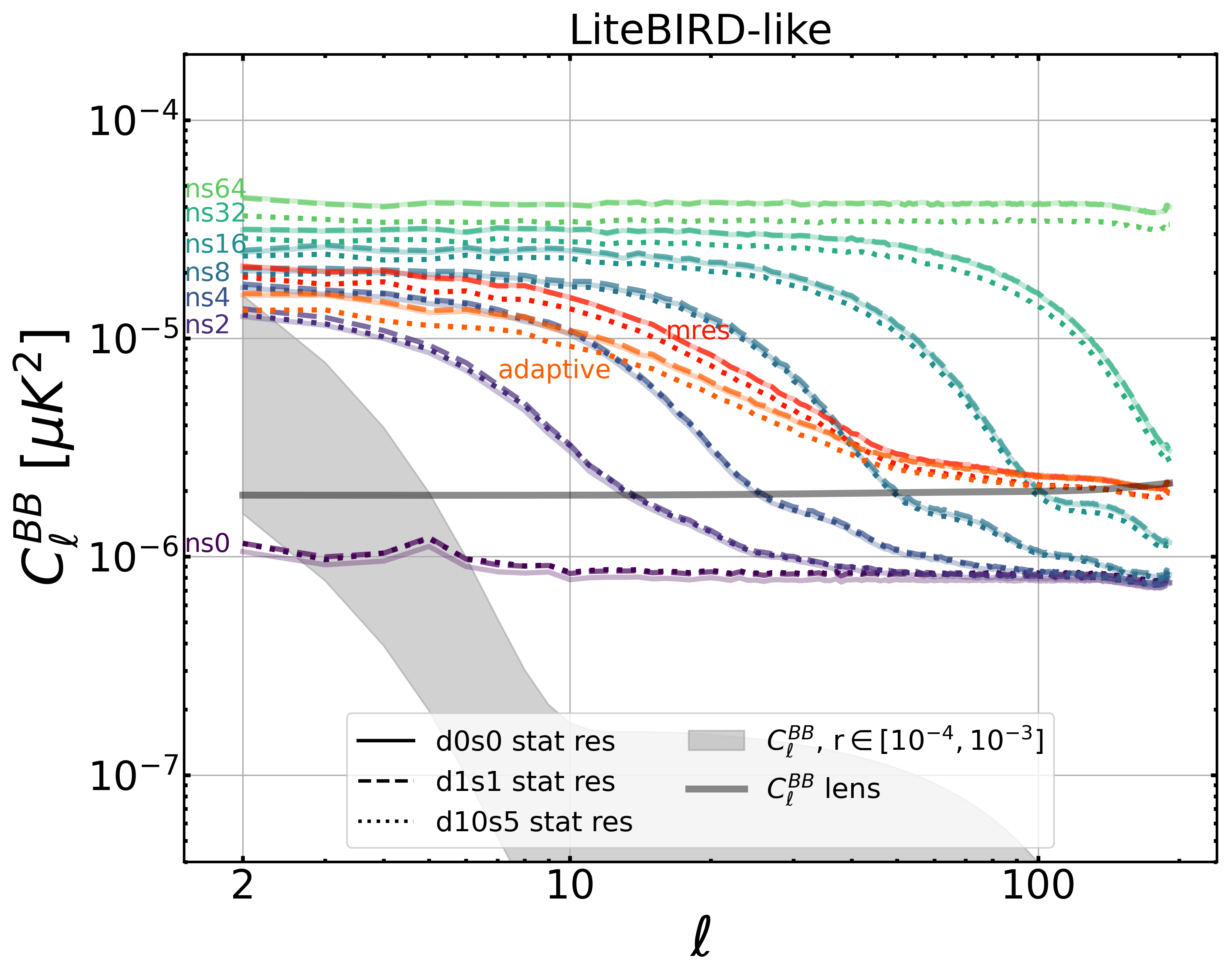}
    \end{minipage}
    \hspace{0.02\textwidth}
    \begin{minipage}[b]{0.48\textwidth}
        \centering
        \includegraphics[width=\linewidth]{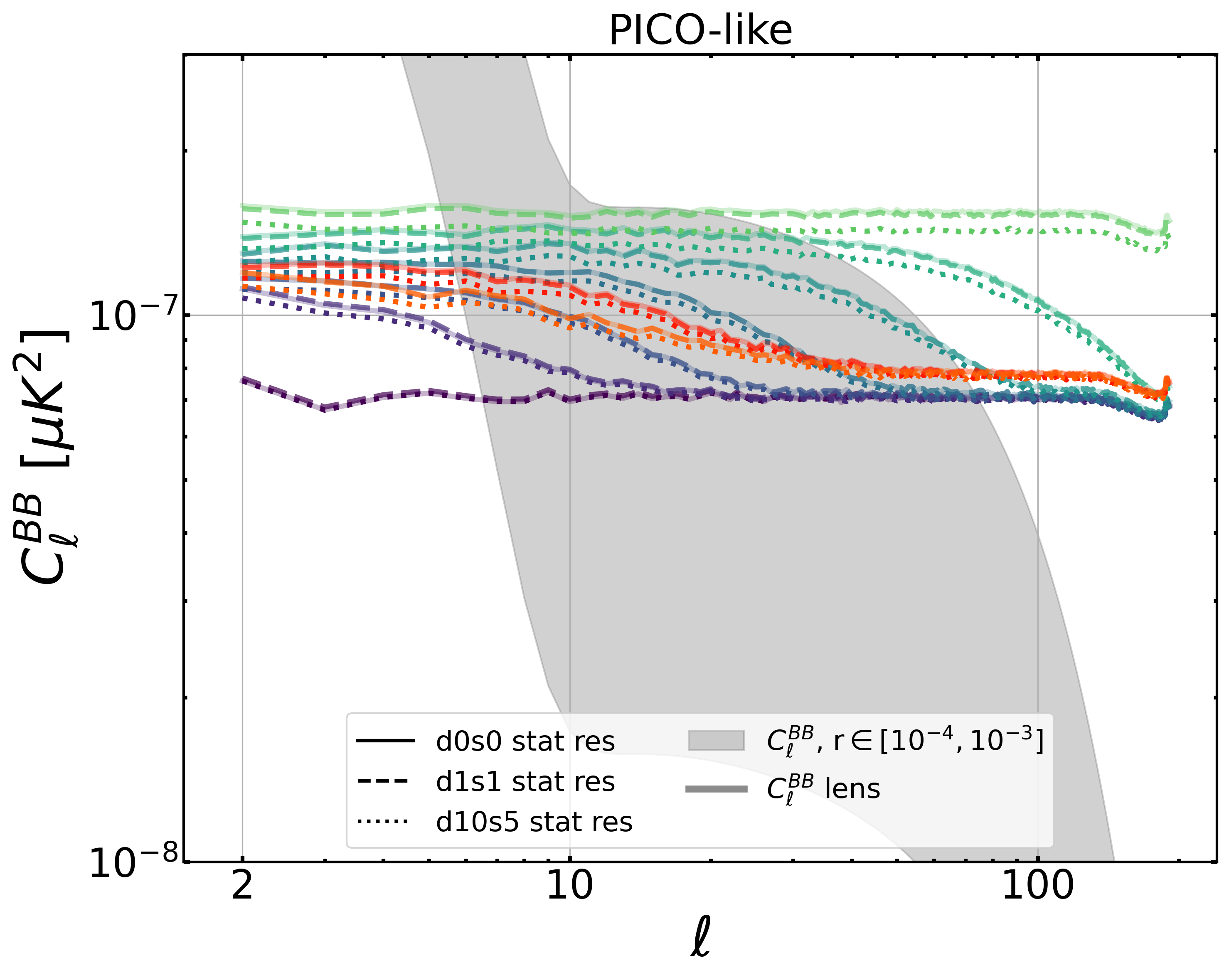}
    \end{minipage}
    \caption{Power spectrum of the statistical foreground residuals for the two missions \textit{LiteBIRD}-like and \textit{PICO}-like, for the same pixel subset configurations of Fig.~\ref{fig:minipage2x2_systematics}, computed with the semi-analytical implementation.
    We show the statistical residuals as computed with the different models studied in this work.
    As expected, we find a mild dependence of the statistical residuals on the foreground model.
    We get to the same conclusion also for the case with ideal clusters, which we do not plot to avoid displaying too many nearly overlapping lines.}
    \label{fig:stat_xforecast_comparing_models}
\end{figure*}

Now that we understand how the semi-analytical framework compares to the simulation-based approach, we can use the semi-analytical implementation to assess the dependence of the statistical estimate of the residuals on the foreground model.
This can be done here in a straightforward way, by directly computing the statistical residual estimates starting from the frequency maps with the different foreground models (hence both the vector $\mathbf{\hat{d}}_{p}$ and $\Bar{\beta}$ in Eqs.~(\ref{eq:residual_map}-\ref{eq:W_1}) are expected to hold slightly different values).
We show those in Fig.~\ref{fig:stat_xforecast_comparing_models}.
We find, both for the LB-like and PICO-like scenarios, larger disagreement with the \texttt{d0s0} in case of the \texttt{d10s5} model, as expected due to both the amplitude template and the spectral parameter templates of this model being more different than the \texttt{d1s1} with respect to the \texttt{d0s0} model.
Note that even if the differences in Fig.~\ref{fig:stat_xforecast_comparing_models} are mild, they do have an impact when estimating the cosmological parameters, in particular in the low-bias cases, where the systematic residuals being low the small mismatch in the statistical residuals can make them higher than the total residuals themselves.
Or, in less extreme cases, there would be an additional bias due to the mis-calibrated estimate of statistical residuals.
This motivates the need to estimate the statistical residuals in a simulation based or realistic data scenario from the simulations/data itself rather than from external simulations.

\begin{figure*}
    \centering
    \begin{minipage}[b]{0.48\textwidth}
        \centering
        \includegraphics[width=\linewidth]{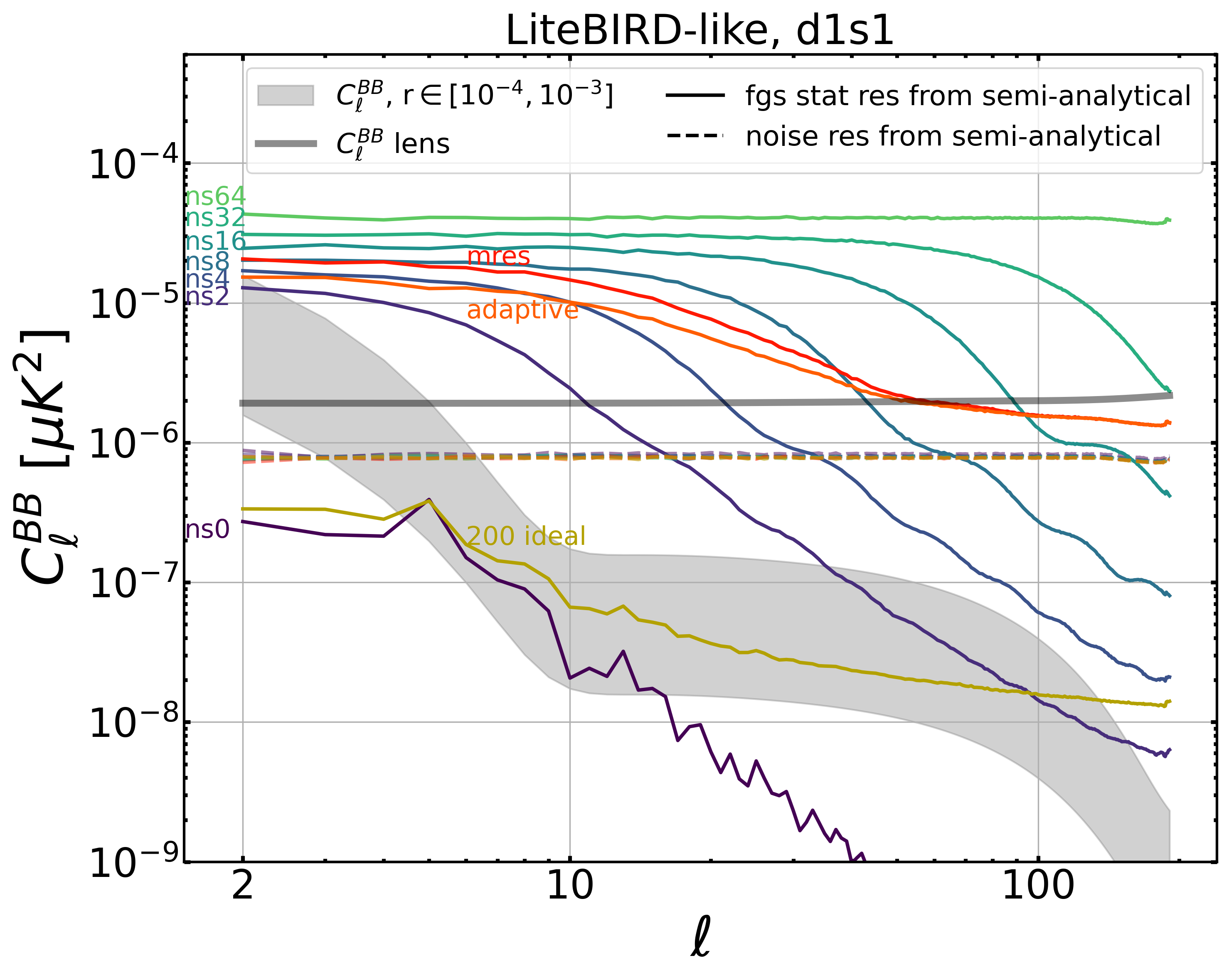}
    \end{minipage}
    \hspace{0.02\textwidth}
    \begin{minipage}[b]{0.48\textwidth}
        \centering
        \includegraphics[width=\linewidth]{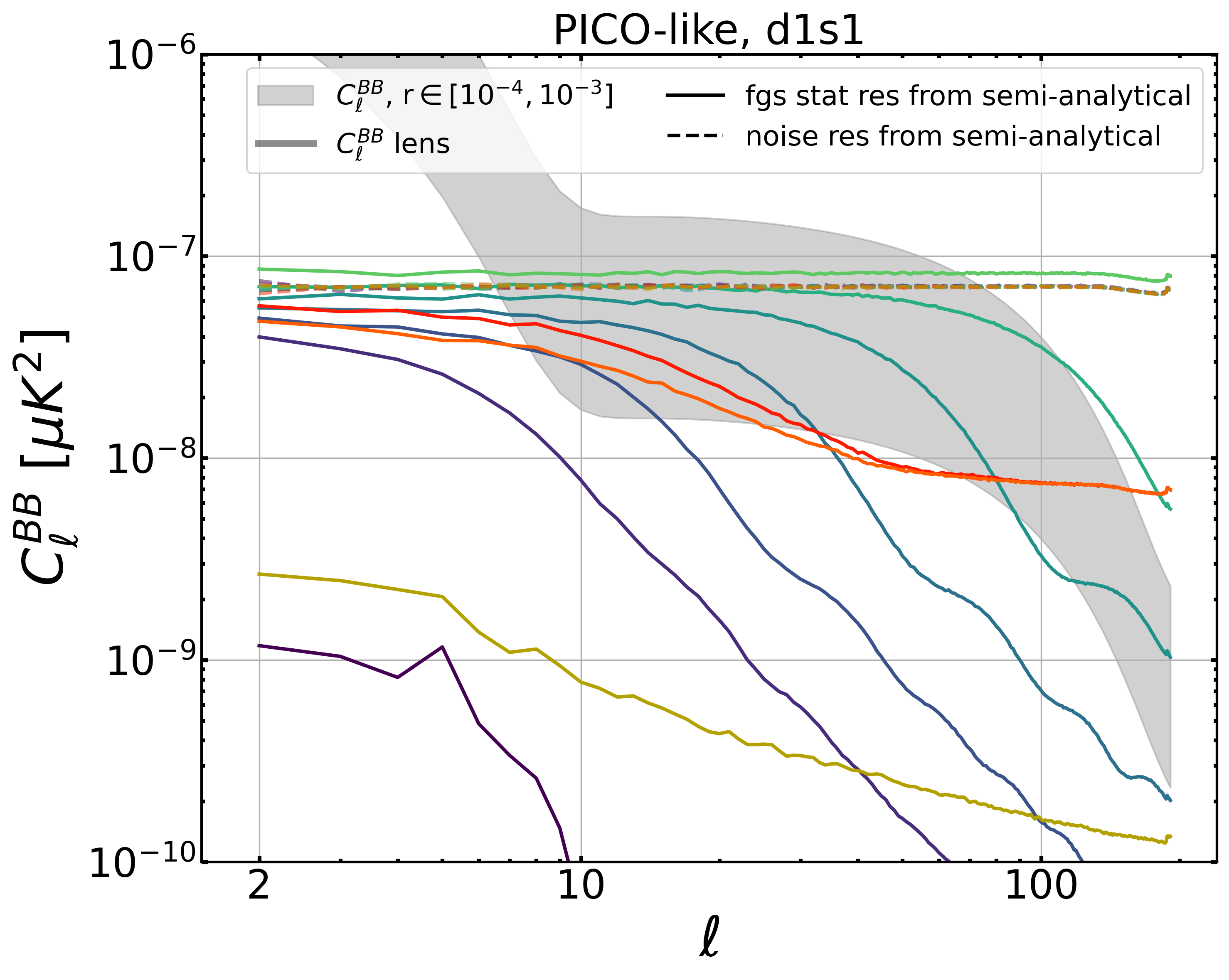}
    \end{minipage}
    \caption{The two contributions to the statistical residuals computed with \texttt{xforecast}, for the LB-like experiment and the PICO-like experiment.
    In both cases the contributions coming from the noise only is only very little dependent on the recovered spectral parameters, giving nearly overlapping lines.
    The solid line is instead the contribution due to the uncertainty on the recovered spectral parameters, and hence dependent on the number of patches and their morphology.}
    \label{fig:stat_noise_fgs}
\end{figure*}

With the forecasting implementation, we can also show the foreground and noise contributions to the statistical residuals independently, see Fig.~\ref{fig:stat_noise_fgs}.
This allows us to gain insights on the shape of the statistical residuals. 
For example, we see that the noise contribution after component separation gives a lower bound for the statistical residuals. 
It is therefore not necessary to decrease the number of pixel subsets beyond the point where the statistical residuals are dominated by the noise.
For the \textsc{HEALPix} patches we can indeed see the statistical foreground contribution behaving as a white spectrum at scales larger than the angular sizes of the patch super-pixels, and becoming a steep red spectrum at angular scales smaller than that.
In particular, for the \texttt{mres} and \texttt{adaptive} cases where the patches for \Bd\ are given by \texttt{nside}=64 patches the foreground contribution to the statistical residuals is the dominating one at any plotted scale.
For what concerns the purely noise contribution to the statistical residuals, which differs in the different cases only for the recovered spectral parameters inserted in Eq.~\eqref{eq:noise_cov_after_comp_sep}, we notice that this dependence is negligible.

\begin{figure}
    \includegraphics[width=\columnwidth]{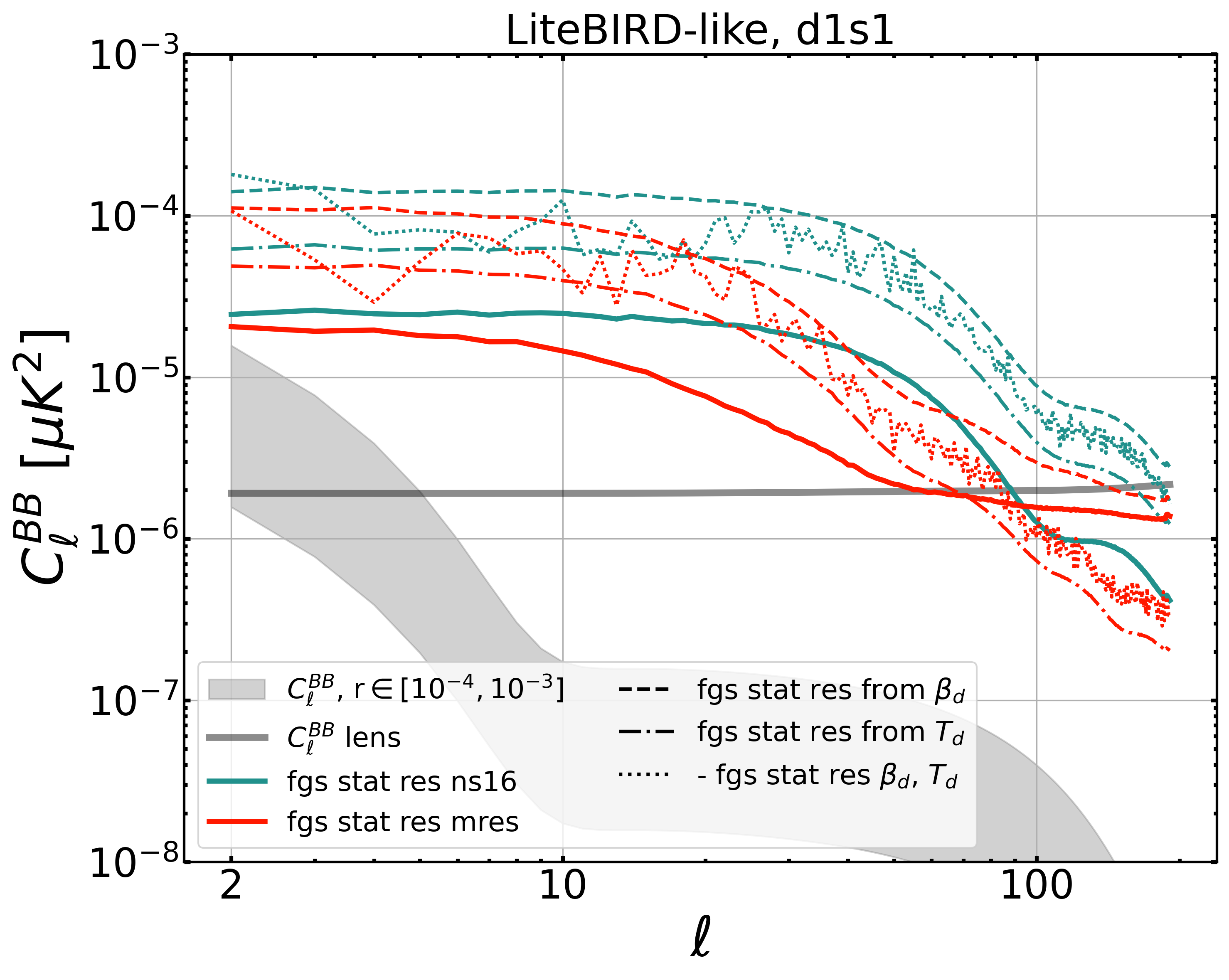}
    \caption{Different contributions to the statistical residuals from \Bd and \Td, for two pixel subset configurations. 
    We show the diagonal spectra of the contributions from these two spectral parameters to the second term of Eq.~\eqref{eq:residual_map}.
    The diagonal contribution for \Bs is instead orders of magnitude lower.
    We also show the only significant cross term, that of \Bd\-\Td.
    Since this term is negative, we plot it as minus itself. 
    This also explains why the individual contributions of \Bd\ and \Td\ are larger than the total statistical residuals.}
    \label{fig:beta_type_contributions_to_stat_res}
\end{figure}

Finally, there is also another decomposition of the statistical residuals that the semi-analytical implementation allows us to do, which is looking at the terms contributing to the foreground statistical residuals for each spectral parameter type.
We usually sum those at the map level, in the second term of Eq.~\eqref{eq:residual_map}, but we can now keep those maps separate and look at the auto-spectrum of each of them and also at their cross-spectra.
This could for example help us elucidate the counter-intuitive relative levels of the statistical residuals in the cases with \texttt{nside}=16 patches and multi resolution with \texttt{nside}=[64,8,4] respectively for [$\beta_\text{d}$, $\mathrm{T}_\text{d}$, $\beta_\text{s}$], as anticipated in the Sect.~\ref{subsection:stat_sim_based}.
A naive explanation could be that in these cases what dominates the statistical residuals is the contribution coming from \Td\ or \Bs, for which the number of patches (and hence the uncertainty on the spectral parameters) decreases going to the multi-resolution configuration.
This can indeed happen in some cases, but looking at the statistical residual contributions from the different spectral parameters, Fig.~\ref{fig:beta_type_contributions_to_stat_res}, we understand that another effect is also occurring in this particular example.
This is the fact that at large scales the anti-correlation of \Bd\ and \Td\ plays an important role, with the residuals due to \Bd\ also significantly reduced and affected by the \texttt{nside}=8 pixelization of \Td. 
We can indeed compute the Pearson correlation coefficient among the maps of statistical residuals of the different spectral parameter types.
We report those values, averaged over the 500 noise realizations and the Q and U maps, in Table~\ref{tab:correlation_coef}.
\begin{table}
\centering
\begin{tabular}{lcc}
\centering
\textbf{Parameters} & \textbf{ns = 16} & \textbf{mres} \\
\hline
\Bd\ , \Td & -0.995 & -0.805 \\
\Bd\ , \Bs\ & 0.376   & 0.162  \\
\Td\ , \Bs\ & -0.351  & -0.176 \\
\hline
\end{tabular}
\caption{Correlation coefficients between \Bd, \Td, and \Bs\ for the \texttt{nside}=16 and multi-resolution cases.}
\label{tab:correlation_coef}
\end{table}

\subsection{Trade-off statistical vs systematics}
\label{subsection:realistic_pixsets}
\begin{figure}
    \centering
    \includegraphics[width=\columnwidth]{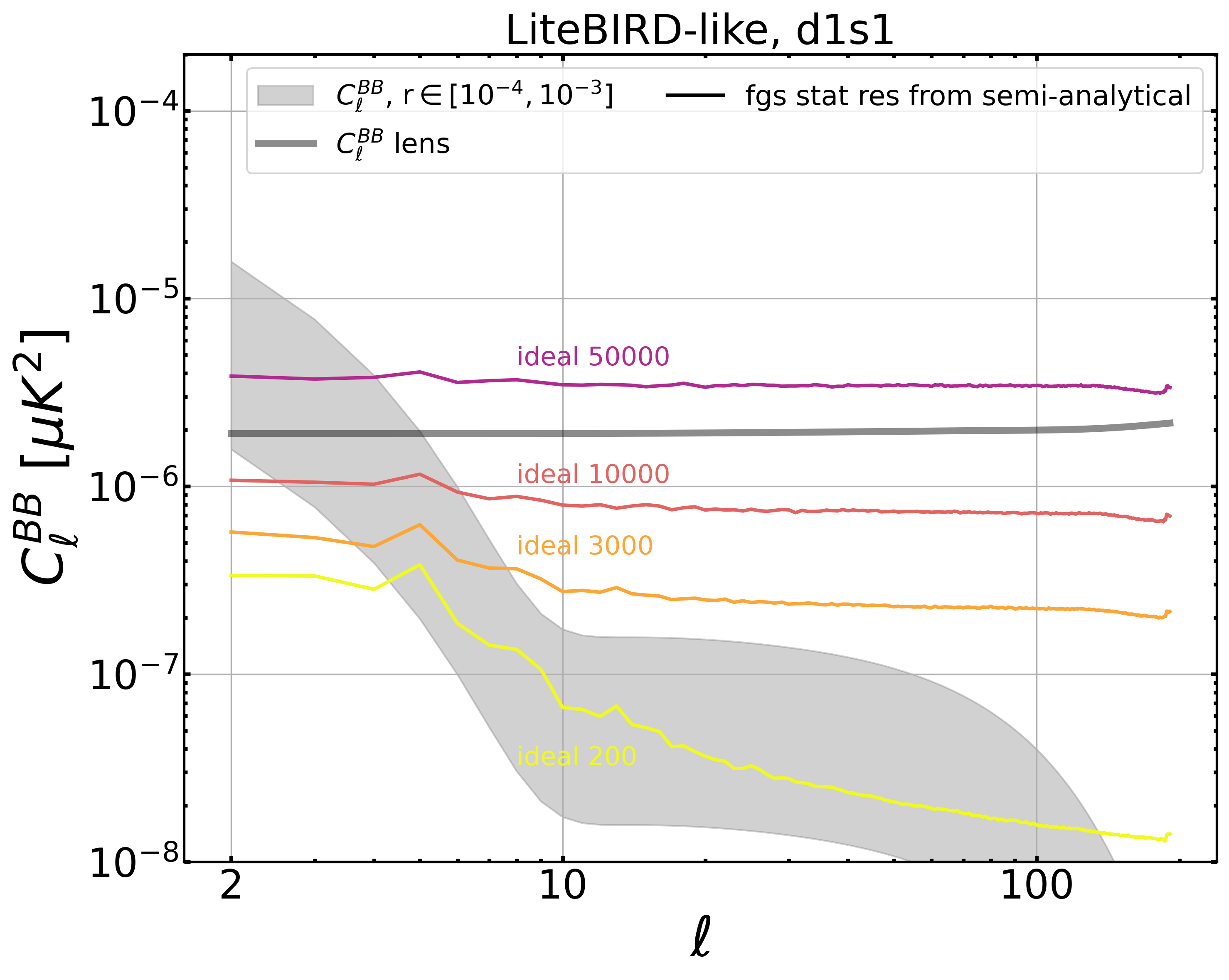}
    \caption{Statistical residuals for different numbers of ideal pixel subsets.
    The numbers on the plots refer to the number of bins in the pixel subsets definition procedure, which, in the 60\% sky area of interest, corresponds to (for the three spectral parameter types \Bd, \Td and \Bs respectively): $[9867, 14484, 13817]$ pixel subsets for the 50000 case; $[2742, 4430, 4391]$ pixel subsets for the 10000 case; $[997, 1598, 1592]$ for the 3000 case; $[95, 147, 126]$ for the 200 bins case.}
    \label{fig:many_clusters}
\end{figure}

We can use the semi-analytical implementation to quickly compute the statistical residuals for the ideal subsets with higher subset numbers.
This is to show that the low statistical residuals in the ideal subset cases in Fig.~\ref{fig:stat_full_vs_xforecast} is not a property of the subset morphology itself, but it is due to the low number of subsets and hence the low uncertainty this gives on the spectral parameters.
The shape of the statistical residuals is however quite different from their shape with \textsc{HEALPix} patches.
Indeed here there is no characteristic size, we can see instead for the lower number of patch cases some large-scale features arising from the large-scale structures of pixel subsets that progressively vanish when increasing the number of patches.
In the latter case the foreground statistical residuals have a white spectrum shape, as the structures in the spectral parameter templates have been mostly broken down in an excessive number of pixel subsets.
We show this in Fig.~\ref{fig:many_clusters} for the LB-like case (analogous considerations could be made for the PICO-like case, but with overall lower statistical residuals due to the different characteristics of the instrument).
We note that the ideal pixel subsets built with the procedure described above at large scales, give lower statistical residuals than the \textsc{HEALPix} cases with the same or, to some extent, even lower number of patches. 
This is because increasing the number of pixel subsets many of those are just made of one or a few pixels, hence only a smaller number of pixel subsets significantly contributes to the statistical residuals.

\begin{figure}
    \centering
    \includegraphics[width=\linewidth]{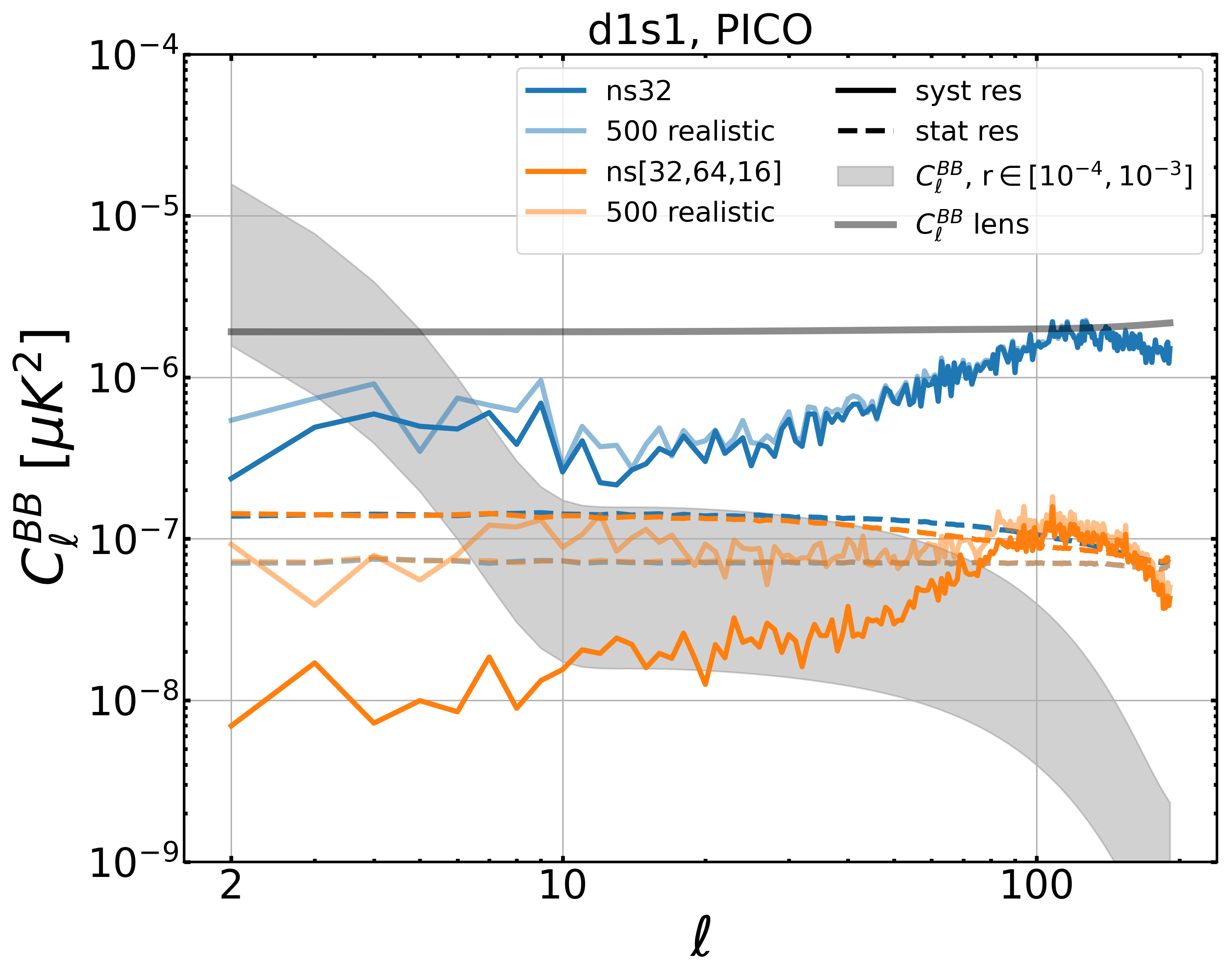}
    \caption{Power spectrum of the systematic and statistical foreground residuals for the \textit{PICO}-like mission, computed with the semi-analytical implementation with a particular \textsc{HEALPix} patches configuration versus with the realistic pixel subsets computed from the recovered spectral parameters from that same configuration.
    In the case of high S/N configurations shown here we recover systematic residuals that are comparable in the two configurations, but we are able to significantly lower the statistical residuals in the case with the pixel subsets which become dominated by the purely noise contribution and almost exactly overlap in the two cases.}
    \label{fig:stat_syst_xforecast_realistic}
\end{figure}

Both $r_{\rm bias}$ and $\sigma(r)$ are sensitive to the degrees of freedom in the component separation. 
Reducing $\sigma(r)$ generally pushes the method to lower the number of parameters, while the systematic residuals typically go down with more numerous parameters.
We introduce a possible way to define pixel subset configurations without using information that would not be accessible with real data (e.g. foreground templates).
The idea is to mimic what is done for the ideal pixel subsets, hence binning the spectral parameter templates.
Without having the true templates of the spectral parameters at our disposal we bin the recovered spectral parameter templates from a component separation with \textsc{HEALPix} patches.
We hence start from an initial component separation with a sufficient number of patches to ensure an acceptable level of systematic residuals, since in the subsequent component separation this level cannot be improved upon.
By binning the recovered spectral parameter maps we then put together the patches with similar values, thus defining the pixel subsets for the second component separation.
In case of low noise levels in the spectral parameter templates binned the found pixel subsets would hence return systematic residuals close to the ones of the first component separation, but statistical residuals lower than that thanks to the lower number of pixel subsets.
We show this (in blue) in Fig.~\ref{fig:stat_syst_xforecast_realistic} for the PICO-like scenario. 
We start from the component separation with \texttt{ns32} and show that we can significantly lower the statistical residuals without significantly impacting the systematic residuals.
However, the presence of noise makes the values of the first spectral parameter templates to be displaced with respect to the true ones, by  different amounts in the different pixels, which further raises the final level of systematic residuals.
This is what happens in the multi-resolution case (\texttt{ns=[32, 64, 16]} for \Bd, \Td, \Bs\ respectively) of Fig.~\ref{fig:stat_syst_xforecast_realistic} .
There, due to the more noisy spectral parameter maps recovered, when building the realistic clusters we are inevitably increasing the systematic residuals, and this remains true unless significantly increasing the number of realistic pixel subsets (which would then make useless the interest of performing this additional component separation).
This approach to define the pixel subsets is hence limited by the bias and the S/N in the recovered spectral parameter templates of the first component separation, hence not bringing a significant improvement in the component separation performance but rather allowing to access different trad-offs of statistical and systematic post-component separation errors.

More generally the interest of reducing the statistical residuals at the expense of the systematic residuals is also justified by the fact that the latter can be further diminished by applying a more aggressive masking of the sky, which has a limit in order not to nullify the benefits of a satellite mission, or by marginalizing, when evaluating the cosmological parameter, over templates of the estimated systematic residual foreground components.

\section{Conclusions}
\label{sec:conclusion}

After having motivated the need for component separation techniques robust against spatially varying foreground emissions, we presented how the parametric map-based methods can address this complexity by performing the component separation independently in different pixel subsets. 
Although all current and future CMB polarization experiments have to deal with spatially varying foreground properties, space missions as full-sky surveys are naturally more sensitive to these. We therefore studied two proposed satellite experiments, namely a \textit{LiteBIRD}-like and a \textit{PICO}-like scenario.
We focused on the parametric component separation method implemented in \fgb, which has been tested and exploited in various contexts in the past, from the instrumental design of existing experiments, the production of forecasts, to the derivation of cosmological results. 
One of the major advantages of this approach over alternative methods is its interpretability, which we explored extensively in a large variety of analysis settings. 
We complemented the simulation-based study of the performance of the method with a semi-analytical modelization of the foreground residuals to move the interpretation even further. This allowed us to derive important results on the performance of the component separation as the hyper-parameters of the analysis vary.
We have shown how systematic and statistical residuals depend on the instrumental configuration, their frequency range and sensitivities.
We have shown that the systematic residuals strongly depend on the foreground model considered, and the statistical residuals are not totally independent on the model either, although the dependence is mild.
The number of pixel subsets appears to be a key hyper-parameter, systematic residuals are reduced as it increases, at the cost of increasing the statistical residuals.
On the other hand, reducing the number of pixel subsets increases the systematic residuals, while the statistical residuals get lower, up to a limit determined by the instrumental sensitivities. 
There is therefore a trade-off to be found between systematic and statistical residuals, which tend to vary in opposite ways. 
However, this trend is impacted by correlations in the foreground parameters: increasing the number of pixel subsets for one parameter can in fact enhance the constraining power on another, as shown with our semi-analytical implementation.
Finally, we discussed how we can aim for more general shapes of the pixel subsets, built without external information, based on the spectral parameter templates recovered from a first
component separation, so reducing the statistical residuals, at the cost of potentially increasing the systematic residuals.
This is only an initial line of investigation, and alternative tracers to build the pixel subsets should also be explored in future works.

We have limited our study to addressing the spatial variability of the foreground models, hence neglecting complexities along the line of sight, which could induce an additional bias on the recovered spectral parameters because of the wrong parametric scaling law choice.
Considering the impact of those in a parametric component separation would be a natural extension of this work.

\section*{Acknowledgments}
The authors thank Radek Stompor for valuable feedback on the final draft.
AR thanks David Alonso for useful discussions during the preparation of this work.
JE acknowledges the SCIPOL project funded by the European Research Council (ERC) under the European Union’s Horizon 2020 research and innovation program (Grant agreement No. 101044073).
The authors acknowledge the use of the \textsc{HEALPix}~\cite{gorski2005healpix} and \texttt{healpy}~\cite{Zonca2019} packages.
Numerical computations were performed on the Glamdring Cluster in the Department of Physics at the University of Oxford.


\bibliographystyle{mnras}
\bibliography{bib}



\appendix

\section{\fgb\ implementation}
\label{ap:implementation}
We describe in the following \fgb\ --- the library implementing the two-step component separation and the forecasting modules described in this work, and whose first version was released in~\cite{2023ascl.soft07021P}.
It was designed to be a flexible, modular and easy-to-use component separation library that can also provide building blocks for tasks and techniques different from those illustrated in this paper. Indeed, it has already been exploited in a series of published articles, for example~\cite{litebird2023probing, ade2019simons, verges2021framework, jost2023characterizing, errard2022constraints}. 

Section~\ref{sec:modules} gives an overview of the structure of the library and the key technical concepts behind the implementation. We invite the reader to check the online documentation for more details, such as the API or a set of pedagogical examples. 
Section \ref{sec:maximization} focuses on what we found to be the most delicate step of the methodology: maximizing the full-sky likelihood fitting thousands of parameters simultaneously.

\subsection{The library}
\label{sec:modules}
\fgb\ is a publicly available python library developed on github\footnote{\url{https://github.com/fgbuster/fgbuster}}. At the time of writing, the package consists of the following modules.

\paragraph{\texttt{separation\_recipes}} is the module that the user typically interfaces with. It provides high-level routines that run component separation on a set of frequency maps given the desired output components. 
It contains routines for multi patch, multi resolution and general shape pixel subsets, internal linear combination etc.
In order to provide easy-to-use interfaces, some assumptions are made (e.g., the inputs are assumed to be \textsc{HEALPix} maps) but custom routines for different use cases can start from these pre-existing functions.

\paragraph{\texttt{component\_model}} contains the most common \texttt{Component}s, objects dedicated to the efficient evaluation of parametric SED models (as well as their derivatives with respect to the parameters). 
New models can be easily constructed from a string containing the analytic expression of the SED.

\paragraph{\texttt{mixing\_matrix}} only stores the object that collects and handles coherently multiple \texttt{Component}s.

\paragraph{\texttt{algebra}} implements all the mathematical operators and equations with no assumptions about the actual origin of the data.
For those that are interested in playing with this set of low-level functions, it is worthwhile highlighting three technical features.
First, all matrices are often block-diagonal, therefore they are passed to functions as \texttt{ndarray}s (tensors) of dimension $(..., n_{\rm row},  n_{\rm col})$ where $n_{\rm row}$ and $n_{\rm col}$ are the number of rows and columns of the \emph{blocks}. ``$...$'' represents any set of extra indices that are necessary to identify the block. 
For example, if the mixing matrix does not mix sky pixels, it will be passed as a tensor with dimensions $(n_{\rm pix}, n_{\rm Stokes}, n_{\rm freq}, n_{\rm comp})$: there is no need to store two sky pixel dimensions.
Second, all operations are compatible with \texttt{numpy} broadcasting. 
To explain it with a practical example, if the mixing matrix is the same for all the Stokes parameters it does not have to be repeated: its dimensions can be $(n_{\rm pix}, 1, n_{\rm freq}, n_{\rm comp})$, thus saving a factor $n_{\rm Stokes}$ in memory space (and computational time, depending on the application).
Reminding that \texttt{numpy} broadcasting aligns dimensions to the right, if the mixing matrix is also constant across the sky it can equivalently have dimensions $(1, 1, n_{\rm freq}, n_{\rm comp})$ or  $(n_{\rm freq}, n_{\rm comp})$.
Finally, all the calculations involving the mixing matrix $\mathbf{A}$ are much faster if this singular value decomposition (SVD) is available: $\mathbf{N}^{-1/2} \mathbf{A} = \mathbf{U} \, {\rm diag} (\mathbf{e}) \mathbf{V}^{\top}$. 
For example, when we fit the spectral parameters we compute $\mathbf{d}^{\top} \mathbf{N}^{-1} \mathbf{A} (\mathbf{A}^{\top} \mathbf{N}^{-1} \mathbf{A})^{-1} \mathbf{A}^{\top} \mathbf{N}^{-1} \mathbf{d}$. 
If the SVD was pre-computed, we get the same results with $|\mathbf{U}^{\top} \mathbf{N}^{-1/2} \mathbf{d}|^2$ (no inversion needed).
This is the reason why most routines taking $\mathbf{A}$ as the input, also have a twin function that takes its SVD instead.

\paragraph{\texttt{cosmology}} gathers several tools that go beyond the component-separated maps. 
In particular, it contains the \texttt{xForecast} tool as described in Sect.~\ref{sec:forecasting}, as well as a simplified version of it only able to deal with pixel independent mixing matrices.

\paragraph{\texttt{observation\_helpers}} contains a few convenient functions for frequent operations (e.g., simulating frequency-maps using \texttt{PySM3} with one line of code).

\subsection{Likelihood maximization}
\label{sec:maximization}
We fit for the linear and non-linear parameters by maximizing the likelihood in Eq.~\eqref{eq:like_pixel_sum}.
In the rest of this section we discuss in detail how we minimize the RHS of Eq.~\eqref{eq:like_pixel_sum}, as the closed-form in Eq.~\eqref{eq:gls} is straightforward to solve, at least for uncorrelated noise covariance.  See e.g.~\cite{poletti2017making, el2022mappraiser} for more involved numerical implementations and exploitation of Eq.~\eqref{eq:gls} in the case of realistic, correlated noise.

\subsubsection{Truncated Newton method}
The minimization has to be compatible with two features of this problem. 
First, we are potentially fitting for thousands of parameters, so the methodology has to scale well.
Second, the full problem cannot be divided into many small ones: if two pixels share the value of a parameter, all the other parameters of these pixels have to be estimated jointly.
We could partition the sky into areas defined by the largest scale in the multi resolution, but we decided not to make any assumption about the morphology in this work to be able to deal with the most general pixel subsets.
Every parameter of the SEDs is associated with an index map, every value of the index map is associated with a region where the index is assumed constant.
Allowing different parameters to have independent index maps is what makes solving the full problem at once necessary.

The implementation accepts any \texttt{scipy} minimizer, however the algorithm we commonly use is a truncated-Newton method \citep{nash1984newton}, implemented in \cite{virtanen2020scipy}, that we now briefly summarize.
Calling $\mathbf{x}$ a vector of $N$ unknowns, Newton methods minimize a function $f(\mathbf{x})$ by taking steps $x_{k+1} - x_k$ along the Newton direction $p_k$ --- itself a solution of the $N \times N$ system ${\nabla^2 f(x_k) p_k = - \nabla f(x_k)}$.
Our sub-class of minimizers settles for an approximate solution to this linear system by performing a limited number of conjugate gradient (CG) iterations (typically $\mathcal{O}(10)$), effectively refining the gradient-descent direction towards Newton's direction.
The Hessian-vector product required by the CG iterations is computed with a finite difference of $\nabla f$ along the direction of the vector --- without ever computing $\nabla^2 f(x_k)$ explicitly.
In summary, a few tens of gradient evaluations are needed for every iteration, plus a few function evaluations for the line search.
The convergence is typically fast, but the number of iterations needed as well as the cost for each of these evaluations depends on the exact nature of the problem, which we now discuss.

\subsubsection{Analytic gradient and Hessian}
First of all, we need efficient tools to evaluate $- \ln \mathcal{L}_{\rm spec}(\{\beta\})$, Eq.~\eqref{eq:like_pixel_sum}, and its gradient.
While some optimizations of our implementation are still certainly possible, the leap that enables the solution of the problem --- even on a single core --- is the implementation of the analytic gradient of the spectral likelihood~\citep{Stompor:2008sf}:
\begin{equation}
   \nabla_\beta \ln \mathcal{L}_{\rm spec} = \sum_p \mathbf{d}_p^\top \, \mathbf{N}^{-1} \, \mathbf{A}_p \left(\mathbf{A}_p^\top \mathbf{N}^{-1} \mathbf{A}_p \right)^{-1} \, \partial_\beta \mathbf{A}_p^\top \, \mathbf{P} \, \mathbf{d}_p,
    \label{eq:grad_like_spec}
\end{equation}
where $\mathbf{P}$ is given by Eq.~\eqref{eq:full_projector}.
For every spectral parameter of the SEDs, we compute $\partial_\beta \mathbf{A}_p$ for all $p$ and thus each element of the sum in Eq.~\eqref{eq:grad_like_spec}.
We then accumulate all the terms of this sum into the appropriate entry of the gradient using the index map of the parameter.
Before implementing Eq.~\eqref{eq:grad_like_spec}, we tried both methods that do not make use of gradients, such as Nelder-Mead~\cite{}, and computing gradients numerically. 
We were unable to reach convergence (within a few hours) even for $\sim 100$ parameters (two spectral parameters at \nside$=2$), while with the analytic gradients we can fit tens of thousands of parameters.

\subsubsection{Convergence}
Defining a good convergence criterion is typically instrumental in achieving quick execution times.
To define when we are close enough to the minimum of $- \ln \mathcal{L}_{\rm spec}$ we can remind that this function is statistically similar to a $\chi^2$ distribution. 
More precisely, $\sum_p \mathbf{d}_p^\top \mathbf{N}^{-1} \mathbf{d}_p - \mathcal{L}_{\rm spec}$ is approximately a $\chi^2$ distributed variable with a number of degrees of freedom equal to the number of measurements in $\mathbf{d}$ minus the total number of parameters fitted for (linear and non-linear).
A natural scale is therefore set by the unity: if for a set of parameters the value of the spectral likelihood is higher than the true minimum by 1, it means that we are squandering a one-sigma constraining power on some combination of the parameters by adding convergence noise on the top of the instrumental one.

Iterations can be stopped, for example, when the decrease of $- \ln \mathcal{L}_{\rm spec}$ from one iteration to the other is less than $\epsilon = 10^{-2}$,
However, in our experience, using much smaller values of $\epsilon$ ($10^{-5}$ or even $10^{-8}$) does not increase too much the total run time.
As emphasized earlier, the convergence process is the most delicate aspect of our analysis from a technical point of view. It certainly deserves to be investigated further both to reduce the time-to-solution and to increase the robustness of the solution. Studying the existence and role of local minima and saddle points (known to be problematic for Newton methods) would be the natural starting point of a future work. 


\bsp	
\label{lastpage}
\end{document}